\documentclass[usegraphicx,usenatbib]{mn2e}

\newcommand{\subs}[1]{$_{\rm #1}$}
\newcommand{\sups}[1]{$^{\rm #1}$}

\newcommand{\BE}{\begin{equation}}
\newcommand{\EE}{\end{equation}}
\newcommand{\kmsn}{km\ s$^{-1}$}
\newcommand{\kmss}{km\ s$^{-1}$ }

\newcommand{\Lsolar}{\mbox{\,$\rm L_{\odot}$}}        
\newcommand{\Msolar}{\mbox{\,$\rm M_{\odot}$}}	      

\newcommand{\vsini}{$\!${\em v\,}sin{\em i}}
\newcommand{\vsinis}{$\!${\em v\,}sin{\em i} }
\def\ga{\mathrel{\hbox{\rlap{\hbox{\lower4pt\hbox{$\sim$}}}\hbox{$>$}}}}
\def\la{\mathrel{\hbox{\rlap{\hbox{\lower4pt\hbox{$\sim$}}}\hbox{$<$}}}}

\title[Chromospheric emission in IC 2391 and IC 2602]
  {The chromospheric emission of solar-type stars in the young open clusters IC 2391 and IC 2602} 
\author[S. C. Marsden, B. D. Carter \& J.-F. Donati] 
  {S. C.~Marsden,$^{1,2}$\thanks{Email: scm@aao.gov.au (SCM); carterb@usq.edu.au (BDC); donati@obs-mip.fr (J-FD)}, B. D.~Carter$^2$\footnotemark[1] and J.-F.~Donati$^3$\footnotemark[1]\\
  $^1$Anglo-Australian Observatory, PO Box 296, Epping, NSW 1710, Australia\\ 
  $^2$Faculty of Sciences, University of Southern Queensland, Toowoomba, QLD 4350, Australia\\ 
  $^3$Laboratoire d'Astrophysique, Observatoire Midi-Pyr\'{e}n\'{e}es, F-31400 Toulouse, France}
\date{}

\pagerange{\pageref{firstpage}--\pageref{lastpage}} \pubyear{2009}

\begin{document}

\label{firstpage}

\maketitle

\begin{abstract}
In this paper we present chromospheric emission levels of the solar-type stars in the young open clusters IC 2391 and IC 2602. High resolution spectroscopic data were obtained for over 50 F, G, and K stars from these clusters over several observing campaigns using the University College London Echelle Spectrograph on the 3.9-m Anglo-Australian Telescope. Unlike older clusters, the majority (28/52) of the solar-type stars in the two clusters are rapid-rotators (\vsinis $>$ 20 \kmsn) with five of the stars being classified as ultra-rapid rotators (\vsinis $>$ 100 \kmsn). The emission levels in the Calcium infrared triplet lines were then used as a measure of the chromospheric activity of the stars. When plotted against Rossby number (N\subs{R}) the star's chromospheric emission levels show a plateau in the emission for Log(N\subs{R}) $\la$ -1.1 indicating chromospheric saturation similar to the coronal saturation seen in previously observed X-ray emission from the same stars. However, unlike the coronal emission, the chromospheric emission of the stars show little evidence of a reduction in emission (i.e.\ supersaturation) for the ultra-rapid rotators in the clusters. Thus we believe that coronal supersaturation is not the result of an overall decrease in magnetic dynamo efficiency for ultra-rapid rotators.
\end{abstract}

\begin{keywords}
stars : rotation -- stars : activity -- stars : late-type -- open clusters and associations: individual: IC 2391, IC 2602
\end{keywords}

\section{Introduction} \label{irtintroduction}

Solar-type stars of spectral types F, G and K arrive on the Zero-Age Main-Sequence (ZAMS) with a wide range of rotation rates, with many stars experiencing extremely rapid rotation (projected rotational velocities, \vsinis values, of 200 \kmss or more). As angular momentum loss is believed to be dependent upon the rotation rate of a star through magnetic braking \citep[e.g.][]{MestelL:1987}, such rapid rotators pose something of a problem as such stars should lose angular momentum much faster than they appear to do.

In order to account for these rapid rotators, dynamo theories of young solar-type stars have used dynamo saturation to slow the angular momentum loss of rapidly rotating stars \citep[e.g.][]{IrwinJ:2007,KrishnamurthiA:1997,BarnesS:1996}. Below a critical rotation rate (the saturation velocity) the strength of a star's magnetic dynamo is believed to be related to the star's rotation rate. However, for stars rotating more rapidly than this it is theorised that the strength of the star's magnetic dynamo is no
longer dependent upon stellar rotation. As a result, the loss rate of angular momentum due to magnetic braking should not increase for solar-type stars with rotation rates greater than the saturation velocity.

The main observational evidence for dynamo saturation comes from the coronal X-ray emission (L\subs{X}/L\subs{bol}, the star's X-ray luminosity divided by the star's bolometric luminosity) of young solar-type stars. When Log(L\subs{X}/L\subs{bol}) is plotted against a measure of rotation rate, such as \vsini, it is seen to initially increase with \vsinis until reaching a value of $\sim$ -3.0 when a plateau in emission level is observed for stars rotating more rapidly than \vsinis $\ga$ 20 \kmsn. Such saturation is also seen in other activity indicators, such as H$\alpha$ and Calcium emission \citep[e.g.][]{SoderblomDR:1993} and this would appear to be consistent with the effect of dynamo saturation \citep{StaufferJR:1997,PattenBM:1996}. However, the mechanism causing this saturation (or even if it is dynamo saturation) remains unknown.

Furthermore, for extremely rapid rotators (\vsinis $\ga$ 100 \kmsn) the Log(L\subs{X}/L\subs{bol}) value of these stars is then seen to drop below the saturation level of Log(L\subs{X}/L\subs{bol}) = -3.0. This effect has been dubbed supersaturation by \citet{ProsserCF:1996b}.

Although the mechanisms causing both saturation and supersaturation are yet to be determined, some suggestions have been put forward, such as centrifugal stripping \citep{JardineM:1999}, and the complete filling of emitting regions \citep*{StepienK:2001}. In order to better constrain both saturation and supersaturation observationally we need to determine what activity indicators show signs of these effects and measure the rotational velocities at which they occur (if at all).

The Calcium II emission of solar-type stars is an excellent diagnostic of chromospheric emission. Traditionally the Ca II H and K resonance lines at 3968\AA\/ and 3934\AA\/ have been used as chromospheric diagnostics and have been employed by \citet{NoyesRW:1984} to look at the relationship between chromospheric emission and rotation rate for solar-type field stars. However observations of the Ca II H and K lines of faint stars can be difficult, due to often low stellar flux in the blue, and the relatively poor blue response of many CCD detectors. In contrast, there have been several studies \citep[e.g.][]{ChmielewskiY:2000,JamesDJ:1997,SoderblomDR:1993,FoingBH:1989,LinskyJL:1979} involving the use of the Calcium II infrared triplet (IRT) lines at 8498\AA, 8542\AA, and 8662\AA, as chromospheric diagnostics. These lines are strong and their location in the red makes them more suitable for CCD observations. The strongest of the IRT lines, the 8542\AA\/ line, is relatively free from blends, and is largely uncontaminated by telluric lines, although \citet{ChmielewskiY:2000} shows that there is some small contamination in the wings of the 8542\AA\/ line, but not the core. The IRT lines have the same upper levels as the Ca II H and K resonance lines, but the lower levels are populated radiatively and are not collisionally controlled like the H and K lines \citep{MallikSV:1997}.

Chromospheric emission in the 8542\AA\/ line has previously been studied by \citet{SoderblomDR:1993} \citep[with the data also presented in][]{KrishnamurthiA:1998} for young solar-type stars in the  Pleiades cluster \citep*[age 130 $\pm$ 20 Myrs;][]{BarradoD:2004} and by \citet{JamesDJ:1997} for the intermediate-age ($\sim$220 Myrs) open cluster NGC 6475. Both results show chromospheric saturation occurring in the rapidly rotating stars with a saturation level in the 8542\AA\/ line of Log(R$^{\prime}$\subs{8542}) $\sim$ -4.2, where R$^{\prime}$\subs{8542} is the flux in the 8542\AA\/ line divided by the star's total bolometric luminosity, see \S~\ref{irtmeasuringemission}. However, due to the lack of ultra-rapid rotators in both these studies, there is no way of telling if the chromospheric emission shows evidence of supersaturation akin to that of the X-ray data. To further the results of \citet{SoderblomDR:1993} and \citet{JamesDJ:1997} this paper determines the chromospheric emission in the Calcium II infrared triplet lines for the solar-type stars in two younger clusters (IC 2391 and IC 2602) containing a higher proportion of rapidly rotating stars, including a number in the supersaturation regime.

\section{The Open Clusters IC 2391 and IC 2602}\label{irtclusters}

The young open clusters IC 2391 and IC 2602 are both southern targets located near the galactic plane. Their HIPPARCOS distances of $\sim$145 pc \citep{vanLeeuwenF:1999} makes them among the closest young open clusters.

The age of the two clusters is given by \citet{StaufferJR:1997} as 30 $\pm$ 5 Myrs. This age is based on both comparisons to evolutionary isochrones as well as the upper main-sequence turnoff age of the clusters. More recent work by \citet{BarradoD:2004} has given an age for IC 2391 of 50 $\pm$ 5 Myrs, based on the lithium depletion of fainter members of the cluster. This has yet to be done for IC 2602.

Given the ages of these two clusters the solar-type stars in IC 2391 and IC 2602 should be on, or in the last phase of evolving to, the ZAMS. At this age, the solar-type stars should not have undergone any significant magnetic braking and thus show a large range of rotational velocities including some stars with ultra-rapid rotation rates (\vsinis $\ga$ 100 \kmsn) placing them within the supersaturation regime. This makes IC 2391 and IC 2602 ideal targets for the study of saturation and supersaturation.

Photometric work on IC 2602 was done by \citet{BraesLLE:1962} and \citet{WhiteoakJB:1961}, while similar observations were undertaken for IC 2391 by \citet{HoggAR:1960}, \citet{BuscombeW:1965}, \citet{PerryCL:1969a}, and \citet{PerryCL:1969b} among others. In the past decade or so there has been renewed interest in these two clusters with ROSAT X-ray observations \citep{RandichS:1995,PattenBM:1996} along with XMM-Newton observations \citep{MarinoA:2005}, as well as a spectroscopic survey by
\citet{StaufferJR:1997} and rotation period determinations for a number of IC 2602 stars by \citet{BarnesSA:1999} and for some IC 2391 stars by \citet{PattenBM:1996}.

The metallicities of the two clusters has been determined by \citet{RandichS:2001a} and \citet{DOraziV:2009} to be close to solar with [Fe/H] = 0.00 $\pm$ 0.01 and [Fe/H] = -0.01 $\pm$ 0.02 for IC 2602 and IC 2391 respectively. 

The designations of the cluster stars used throughout this paper are as follows. For IC 2391, VXR: designation from \citet{PattenBM:1996}, H: from \citet{HoggAR:1960}, L: from \citet{LyngaG:1961}, and SHJM: from \citet{StaufferJR:1989}. For IC 2602, R: designation from \citet{RandichS:1995}, B: from \citet{BraesLLE:1962}, and W: from \citet{WhiteoakJB:1961}. Many of the stars have multiple designations and where possible the designations from \citet{PattenBM:1996} and \citet{RandichS:1995} have been used for the stars in IC 2391 and IC 2602 respectively.

For stars in IC 2391 the quoted photometry is from \citet{PattenBM:1996}, \citet{LyngaG:1961}, \citet{HoggAR:1960}, and \citet{StaufferJR:1989}. The rotational periods are from \citet{PattenBM:1996} and the previous \vsinis values are from \citet{StaufferJR:1989} and \citet{StaufferJR:1997}. For IC 2602 the photometry is taken from \citet*{ProsserCF:1996a}, \citet{RandichS:1995}, and \citet{BarnesSA:1999}. The rotational periods are from \citet{BarnesSA:1999} and the previous \vsinis values are from \citet{StaufferJR:1997}.

Because of the similarities between the two clusters, in age, distance, metallicity, etc., and the relative paucity of stars in the clusters (compared to say the Pleiades), IC 2391 and IC 2602 are treated as one cluster. We present the results from both clusters together, although stars from each cluster are represented by different symbols.

Most of the information, such as names, membership, positions, colours, etc., for the stars in the two clusters can be found in the WEBDA database\footnote{WEBDA database - http://www.univie.ac.at/webda/}.

\section{Selection of Targets}\label{irtselection}

Potential members of IC 2602 were identified by \citet{RandichS:1995} from ROSAT X-ray data, with a number of active low-mass candidate members of the cluster found. A similar approach was taken by \citet{PattenBM:1996} for the members of IC 2391. \citet{ProsserCF:1996b} and \citet{StaufferJR:1997} then carried out photometric and spectroscopic confirmation of membership studies (along with \vsinis measurements), and a list of active low-mass members of the clusters was produced. The listing that \citet{StaufferJR:1997} gives for IC 2391 is not as complete as that given by \citet{PattenBM:1996} so we have added some of the \citet{PattenBM:1996} stars to the list of potential members. This list is still not a complete census of the solar-type members in the two clusters, because, as \citet{StaufferJR:1997} points out, in neither cluster does the ROSAT survey used to determine membership cover the entire area of the sky over which cluster members are likely to occur. \citet{StaufferJR:1997} argue that although the list is incomplete it should not be overly biased (towards active stars) by the selection process (using X-ray emission to determine membership) for at least the G dwarfs in the two clusters. For the K dwarfs they believe that any bias should not be a large effect. A more comprehensive explanation of their selection process along with estimates of the completeness of the sample is given in \citet{StaufferJR:1997}.

To the \citet{StaufferJR:1997} list (along with additional stars from \citet{PattenBM:1996}), we have added further stars from the \citet{BraesLLE:1962} and \citet{WhiteoakJB:1961} photometric studies of IC 2602, as well as some IC 2391 stars from \citet{HoggAR:1960}, \citet{LyngaG:1961}, and \citet{StaufferJR:1989}, again not observed by \citet{StaufferJR:1997}. The IC 2602 photometric studies by \citet{BraesLLE:1962} and \citet{WhiteoakJB:1961} only went as deep as V $\sim$ 11, meaning that stars added to the list from these studies were predominantly late-F/early-G stars.

With the possible exception of the above mentioned additional stars, we believe that most of the stars in our target list should be cluster members. However a further membership determination was made based on radial velocity and Lithium line strength, see \S~\ref{irtmembership}.

For this study as many of the solar-type stars in our target list were observed as possible in the time available, without biasing the observations through colour selection effects. This was achieved with  observations of all of the selected stars in both IC 2391 and IC 2602 with 0.4 $\le$ (V-I\subs{C})\subs{0} $\le$ 1.4, corresponding to $\sim$ early-F to $\sim$ mid-K stars. Due to time constraints and the wish not to bias the observations by only obtaining data on a few lower-mass stars, all stars with (V-I\subs{C})\subs{0} $\ga$ 1.4 were excluded. Thus the observations should constitute a relatively unbiased sample of solar-type stars from early-F through to mid-K. In IC 2391 32 stars were observed from the target list, while in IC 2602 51 stars were observed from the target list.

A list of the stars with 0.4 $\le$ (V-I\subs{C})\subs{0} $\le$ 1.4 observed in IC 2391 and IC 2602 are given in Tables~\ref{irt2391rot} and~\ref{irt2602rot} respectively.

\begin{table*}
\caption{}
\vbox to 220mm{\vfil Landscape Table 1 to go here. 
\vfil}
\label{irt2391rot}
\end{table*}

\begin{table*}
\caption{}
\vbox to 220mm{\vfil Landscape Table 2 to go here.
\vfil}
\label{irt2602rot}
\end{table*}

\section{Observations}\label{irtobservations}

The spectroscopic data for this project were collected over 3 separate observing campaigns at the 3.9-m Anglo-Australian Telescope (AAT) located at Siding Spring Observatory in New South Wales, Australia. The 3 observing runs consisted of 3, 3, and 2 nights in March 2000, January 2001, and February 2001, respectively.

The spectroscopic data were obtained using the University College London Echelle Spectrograph (UCLES). The detector used for the first two runs (March 2000 and January 2001) was the Deep Depletion MITLL3 CCD with 2048 $\times$ 4096 15 $\mu$m square pixels. This chip was chosen for its excellent red response and low fringing. For the third run (February 2001), the chip used was the lumogen coated MITLL2A CCD, again with 15 $\mu$m square pixels. This chip does not have the improved red response of the MITLL3, however the MITLL3 chip was unavailable at the time.

Since both of the chips are larger than the unvignetted field of the UCLES camera, a smaller window format (2048 $\times$ 2896 pixels) was used to reduce read out time. Using the 31.6 gr/mm grating, 57 orders (\#55 to \#111) could be fitted onto the detector window. Only 49 (\#63 to \#111) were reduced (the remaining orders had little or no signal in them) giving full wavelength coverage from $\sim$5000\AA\/ to $\sim$9000\AA. While the actual wavelength range varied slightly from run to run, this variation was minimal.

With a slit-width of $\sim$1 arcsecond the spectral resolution varied from $\sim$44,000 to $\sim$46,000 during the 3 runs, giving a velocity resolution of $\sim$6.8 \kmss to $\sim$6.5 \kmsn.

Several exposures were taken consecutively for each star and added together to improve the signal-to-noise and also in case cosmic rays affected the IRT lines. The raw frames were then reduced and converted into wavelength calibrated spectra using the ESpRIT (Echelle Spectra Reduction: an Interactive Tool) optimal extraction routines of \citet{DonatiJF:1997b}.

\section{Results}\label{irtresults}

\subsection{Measuring Rotational and Radial Velocities \label{irtmeasuringvelocity}}

The major part of this study involves investigating the relationship between chromospheric activity and the rotation rates of solar-type stars in IC 2391 and IC 2602. Although many of the stars in the observation list have previous \vsinis values measured by \citet{StaufferJR:1989,StaufferJR:1997}, a number of stars still had none. To have a consistent \vsinis determination for the entire observation list, it was decided to re-determine the values of \vsinis of all the observed stars. In order to do this the Least-Squares Deconvolution (LSD) profiles for the stars were extracted from the spectrum of each star using the routines of \citet{DonatiJF:1997b}.

An LSD profile can be considered as the resultant sum of the several thousand photospheric lines contained within each Echelle spectrum, yielding an ``average'' single high S/N profile (see \citealt{DonatiJF:1997b} for more details). The resultant gain in S/N from this process was $\sim$10 times that of the S/N in the 8542\AA\/ line, depending on the spectral type of the star observed. The LSD profiles of three low activity stars of differing spectral types were also produced from the observations. These stars were HD 16673, $\alpha$ Cen A, and $\alpha$ Cen B, with spectral types of F6V, G2V and K1V respectively.

\begin{figure}
	\begin{center}
	\includegraphics[angle=90,width=8.4cm]{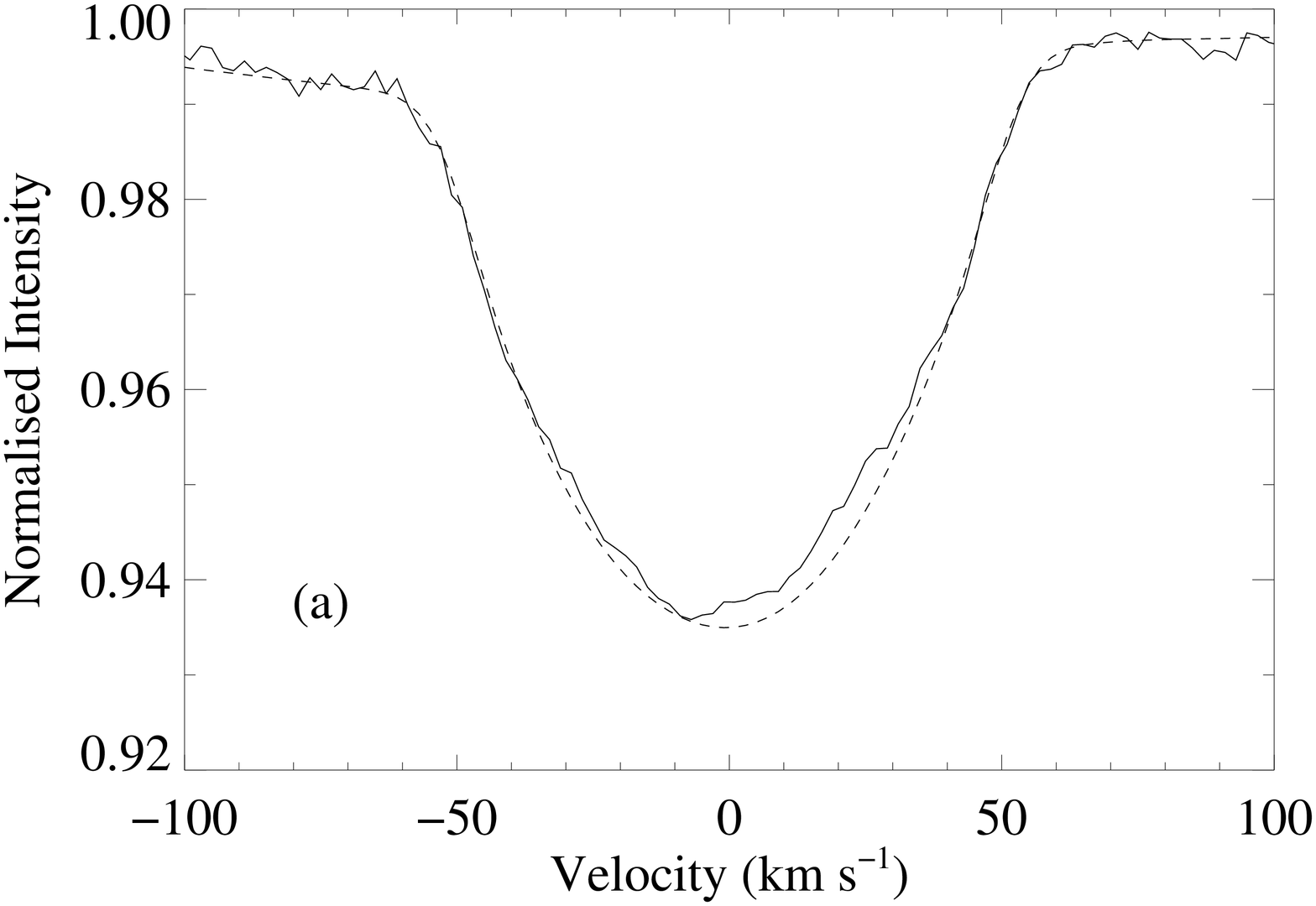}
	\includegraphics[angle=90,width=8.4cm]{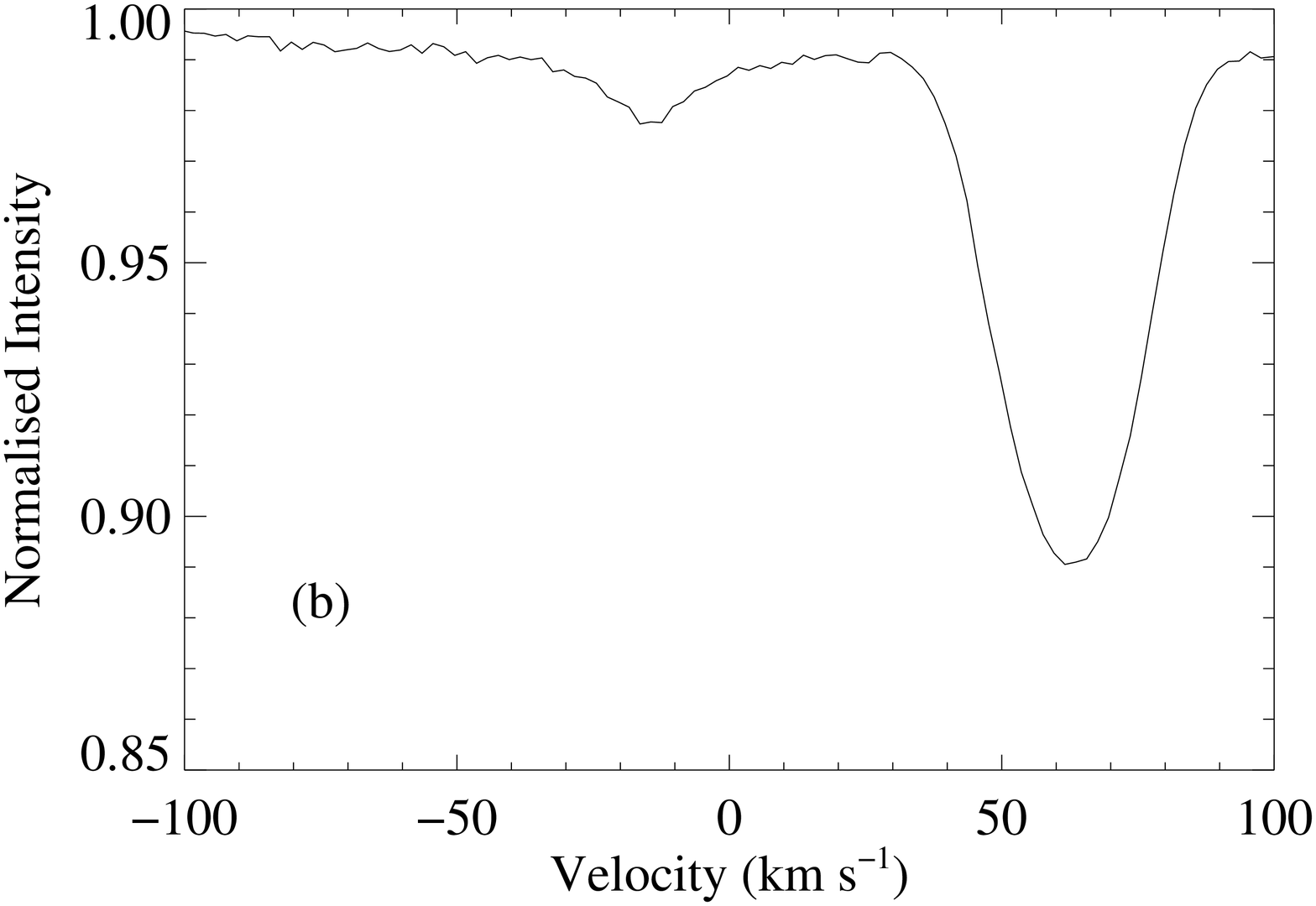}
	\caption{LSD profiles for (a) VXR66 (from IC 2391) and (b) R46 (from IC 2602). The profiles have been zeroed to (a) the radial velocity of VXR66, 15 \kmss and (b) the average cluster velocity of IC 2602, 17.4 \kmsn. The dashed line in (a) shows the LSD profile of the inactive comparison star HD16673, rotationally broadened to match the \vsinis of VXR66 (\vsinis = 52 \kmsn). The asymmetric bumps in the bottom of the LSD profile of VXR66 are most likely due to spots on the star's photosphere. R46 is most likely a double-lined spectroscopic binary.}
	\label{irtlsd}
	\end{center}
\end{figure}

Based on the colour of the target star an inactive star of similar spectral type was then rotationally broadened to match the target star's LSD profile. Given the high S/N of the LSD profiles (from $\sim$300 to $\sim$900) the fit to the rotationally broadened profile could usually be found to within $\pm$1 \kmsn. This error is based on internal consistency within the method by observing the same star over several epochs. The \vsinis values of $\alpha$ Cen A and $\alpha$ Cen B are given by \citet{ValentiJA:2005} as 2.3 \kmss and 0.9 \kmss respectively. Given the instrumental resolution of the observations (6.5  to 6.8 \kmsn) we estimate that the lower \vsinis limit we could determine for our stars was $\sim$7 \kmss when using $\alpha$ Cen A or B as a template. This is not the case for HD 16673. The \vsinis value for this star is given by \citet{NordstromB:2004} as 8 \kmsn. This means that for those stars using HD 16673 as a template the lower \vsinis limit was 11 \kmsn. However, no slowly-rotating stars determined to be cluster members used HD 16673 as a temple star. The LSD profile of one of the target stars and the fit to a rotationally broadened low activity star is given in Figure~\ref{irtlsd}(a). A list of the \vsinis values for the observed stars, determined using this LSD method (along with comparisons to previous measurements) is given in Tables~\ref{irt2391rot} and~\ref{irt2602rot}.

\begin{figure}
	\begin{center}
	\includegraphics[angle=90,width=8.4cm]{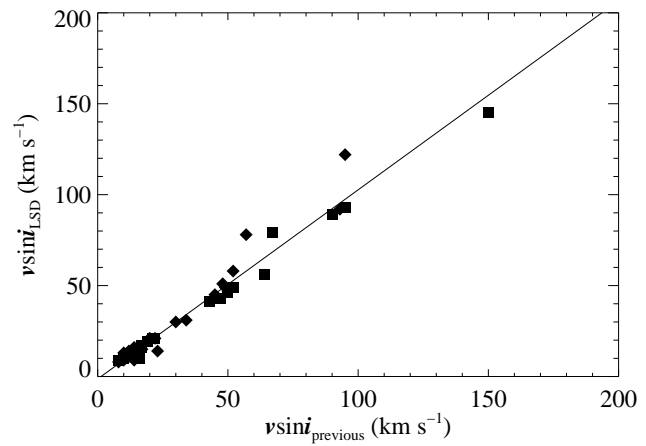}
	\caption{Comparison of \vsinis measurements for the stars in IC 2391 and IC 2602 (excluding those stars with \vsinis $\le$ 7 \kmsn). Squares represent stars from IC 2391 and diamonds IC 2602. The solid line represents the linear fit to the data given in Equation~\ref{irtvsinirel}.}
	\label{irtvsinicomp}
	\end{center}
\end{figure}

Using those single stars from Tables~\ref{irt2391rot} and~\ref{irt2602rot} that have previously had their \vsinis values determined, the relationship between the \vsinis measured here using the LSD profiles and previous measurements \citep[from][]{StaufferJR:1989,StaufferJR:1997} was determined. This relationship is shown in Figure~\ref{irtvsinicomp} and using a linear fit was found to be (with 1 $\sigma$ errors): 
\BE
v\,{\rm sin}i_{\rm LSD} {\rm = (1.04 \pm 0.03)} \times v\,{\rm sin}i_{\rm previous} {\rm - (1.34 \pm 1.63)}. \label{irtvsinirel} 
\EE

Along with being a useful technique for the measurement of rotational broadening, the high S/N of the LSD profile also makes it a good way to determine if a star is a double-lined spectroscopic binary. Tables~\ref{irt2391rot} and~\ref{irt2602rot} also note those stars suspected of being double-lined binaries based on their LSD profiles. The LSD profile of one of these binaries is given as an example in Figure~\ref{irtlsd}(b).

All single members of a cluster are expected to have a similar radial velocity. During the fitting of the LSD profiles of the inactive stars to the target stars in IC 2391 and IC 2602 we have also determined the radial velocity of our target stars. These are heliocentric radial velocities which have then been shifted to match the centroid of the LSD profile of the telluric lines in each spectrum to account for any changes in temperature or pressure experienced by the spectrograph during the coarse of the observations. This should give a zero point error of $\sim\pm$0.3 \kmss for both the target and inactive stars. Thus given an estimated error of  $\sim\pm$1 \kmss for the method we estimate that the radial velocity measurements are accurate to $\sim\pm$1.5 \kmsn. The radial velocities of the observed stars measured through this technique are again given in Tables~\ref{irt2391rot} and~\ref{irt2602rot}. The mean and standard deviation of the radial velocities for the two clusters (measured from those stars given a M membership classification, see \S~\ref{irtmembership}) was found to be 15.9 $\pm$ 0.7 \kmss and 17.4 $\pm$ 1.0 \kmss for IC 2391 and IC 2602 respectively.

\subsection{Cluster Membership \label{irtmembership}}

As mentioned in \S~\ref{irtselection}, we believe that most of the stars in out target list should in fact be cluster members, based on previous selections. However, we have made a further determination of the cluster membership based on the strength of the star's Lithium 6708\AA\/ line in our observations and our determination of the radial velocity of the star. The strength of the Lithium line was only determined approximately with respect to the strength of the nearby Ca 6717\AA\/ line, and given as Strong, Weak, or None. A more detailed study of the Lithium in the two clusters has been undertaken by \citet{RandichS:1997a}, \citet{RandichS:2001b}, and \citet{RandichS:2001a}. 

Most of the single stars in Tables~\ref{irt2391rot} and~\ref{irt2602rot} with strong Lithium appear to have radial velocities around 16 \kmss for IC 2391 and 17 \kmss for IC 2602. Thus in order to qualify as a member of the cluster (M classification) stars have to be single (i.e. no evidence of a binary in their LSD profile), have strong Lithium, and have a radial velocity of 16 $\pm$ 2 \kmss for IC 2391 stars and 17 $\pm$ 2 \kmss for IC 2602. Stars which are classified as probable members (denoted by a ? classification) have a strong Lithium feature and are either binary or if single have a radial velocity of 16 $\pm$ 9 \kmss for IC 2391 stars or 17 $\pm$ 9 \kmss for IC 2602 stars. All other stars are classified as non-members (N classification). This membership is given in Tables~\ref{irt2391rot} and~\ref{irt2602rot}. In our analysis of the chromospheric emission of the stars (\S~\ref{irtmeasuringemission}) we have included all single stars with either M or ? classifications.

Since these observations there have been other determinations of cluster membership, such as that by \citet{PlataisI:2007} for IC 2391. The membership list of \citet{PlataisI:2007} contains a number of new IC 2391 members not contained in our list, however this was published too late for our study. The \citet{PlataisI:2007} list also has membership determinations for the stars in our list. The determination of membership agrees reasonably well with our list. In addition, \citet{PlataisI:2007} list several IC 2391 stars as single-lined spectroscopic binaries that we treat as single,  however, the contribution from the secondary to the chromospheric emission should be minimal so we have included them. Table~\ref{irtexceptions}  gives a list of differences between our designations and that of \citet{PlataisI:2007}. \citet{StaufferJR:1997} list the star R80 (IC 2602) as a possible non-member. Based on our measurement of its radial velocity and Lithium we feel it is consistent with membership,  and hence we have included it as such.

\begin{table}
\begin{center}
\caption{List of differences in determination of cluster membership and binarity between our work and \citet{PlataisI:2007} for the solar-type stars IC 2391. M: cluster member, N:,  non-member, ?: probable member, B: Binary.}
\label{irtexceptions}
\begin{tabular}{lcc}
\hline
Star & Our & \citet{PlataisI:2007} \\
Name & designation           & designation\\
\hline
L32 & M & M + B\\ 
VXR02B & N & M\\
VXR05 & ? + B & M + B\\
VXR07 & ? & M +  B\\
VXR08 & N & M + B\\
VXR11 & ? + B & M + B\\
VXR30 & N & M + B\\
VXR31  & ? & N\\
VXR35A & M & ?\\
VXR44 & ? & M + B\\
VXR45A & M & ?\\
VXR50A & ? & N\\
VXR52 & M & M + B\\
VXR62A & ? & M\\
VXR67A & ? & M + B\\
VXR70 & ? & M\\
VXR78  & ? & N\\
\hline
\end{tabular}
\end{center}
\end{table}

\subsection{Stellar Parameters \label{irtparameters}}

The effective temperatures of the stars given in Tables~\ref{irt2391rot} and~\ref{irt2602rot} were calculated from (V-I\subs{C})\subs{0} colours and the colour vs.\ T\subs{eff} relationships of \citet*{BessellMS:1998}. If no (V-I\subs{C})\subs{0} measurement was available for the star then (B-V)\subs{0} was used. The (V-I\subs{C}) and (B-V) values were converted to unreddened values by using the reddening parameters for the clusters. E(B-V) = 0.04 for IC 2602 from \citet{BraesLLE:1961} and E(B-V) = 0.006 (rounded to 0.01) for IC 2391 from \citet{PattenBM:1996}. The E(V-I\subs{C}) values for the two clusters were found by multiplying the E(B-V) values by 1.25 as given in \citet{PinsonneaultMH:1998}, thus giving values of E(V-I\subs{C}) = 0.05 and E(V-I\subs{C}) = 0.0075 (rounded to 0.01) for IC 2602 and IC 2391 respectively.

The luminosities of the target stars were calculated from:
\BE
{\rm M_{bol} = BC_{I} + I = 4.74 - 2.5log(\frac{L}{\Lsolar})}
\label{irtmbol} 
\EE
where 4.74 is the M\subs{bol} for the Sun \citep{BessellMS:1998} and BC\subs{I} is the bolometric correction in the I band calculated using \citet{BessellMS:1993}. The distance moduli used for the clusters were 5.82 and 5.81 for IC 2391 and IC 2602 respectively taken from the HIPPARCOS distances of \citet{vanLeeuwenF:1999}. If no (V-I\subs{C})\subs{0} value was available then:
\BE
{\rm M_{bol} = BC_{V} + V}
\EE
was used instead, with the BC\subs{V} calculated from \citet{BessellMS:1998}.

\begin{figure}
	\begin{center}
	\includegraphics[width=8.4cm]{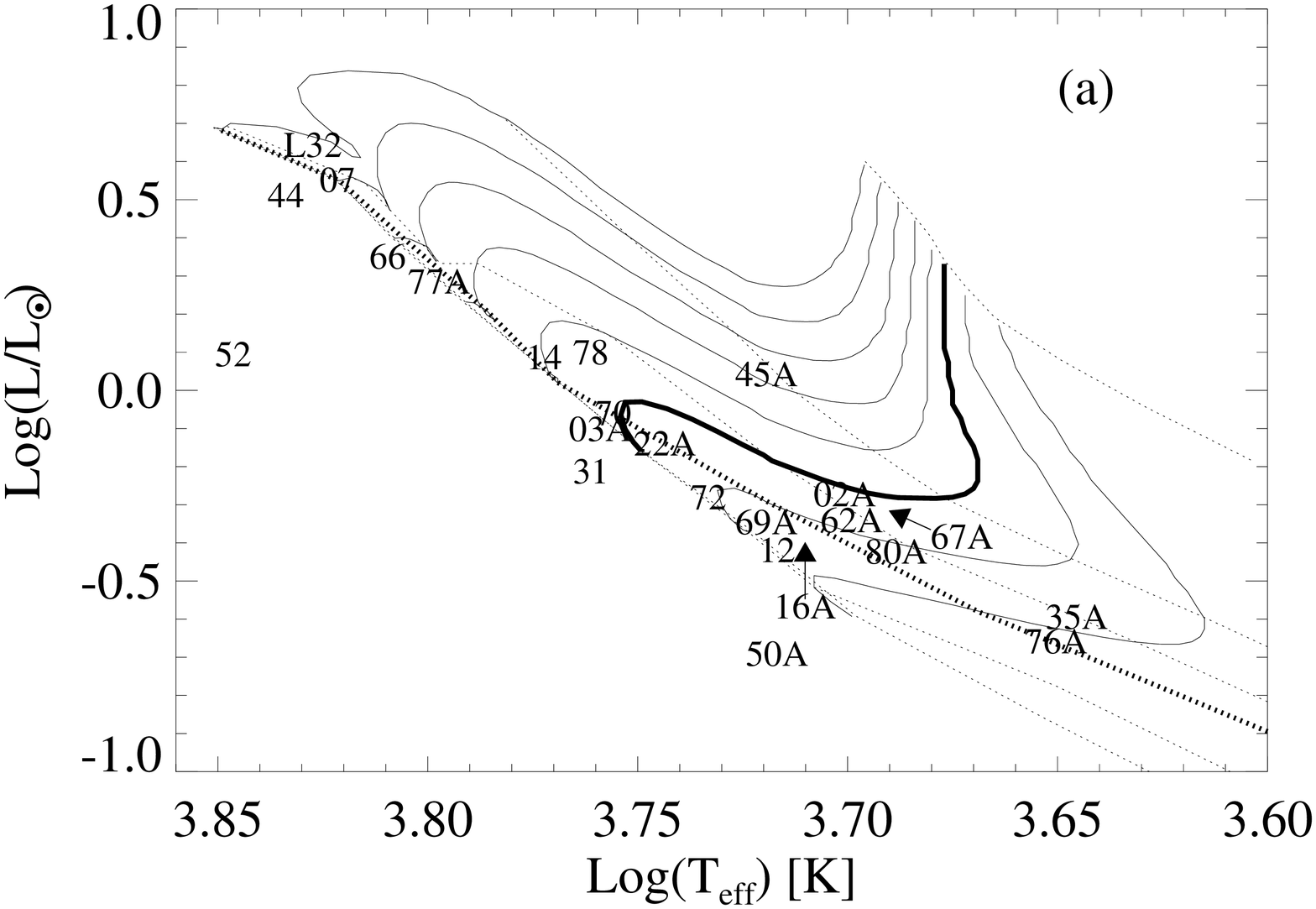}
	\includegraphics[width=8.4cm]{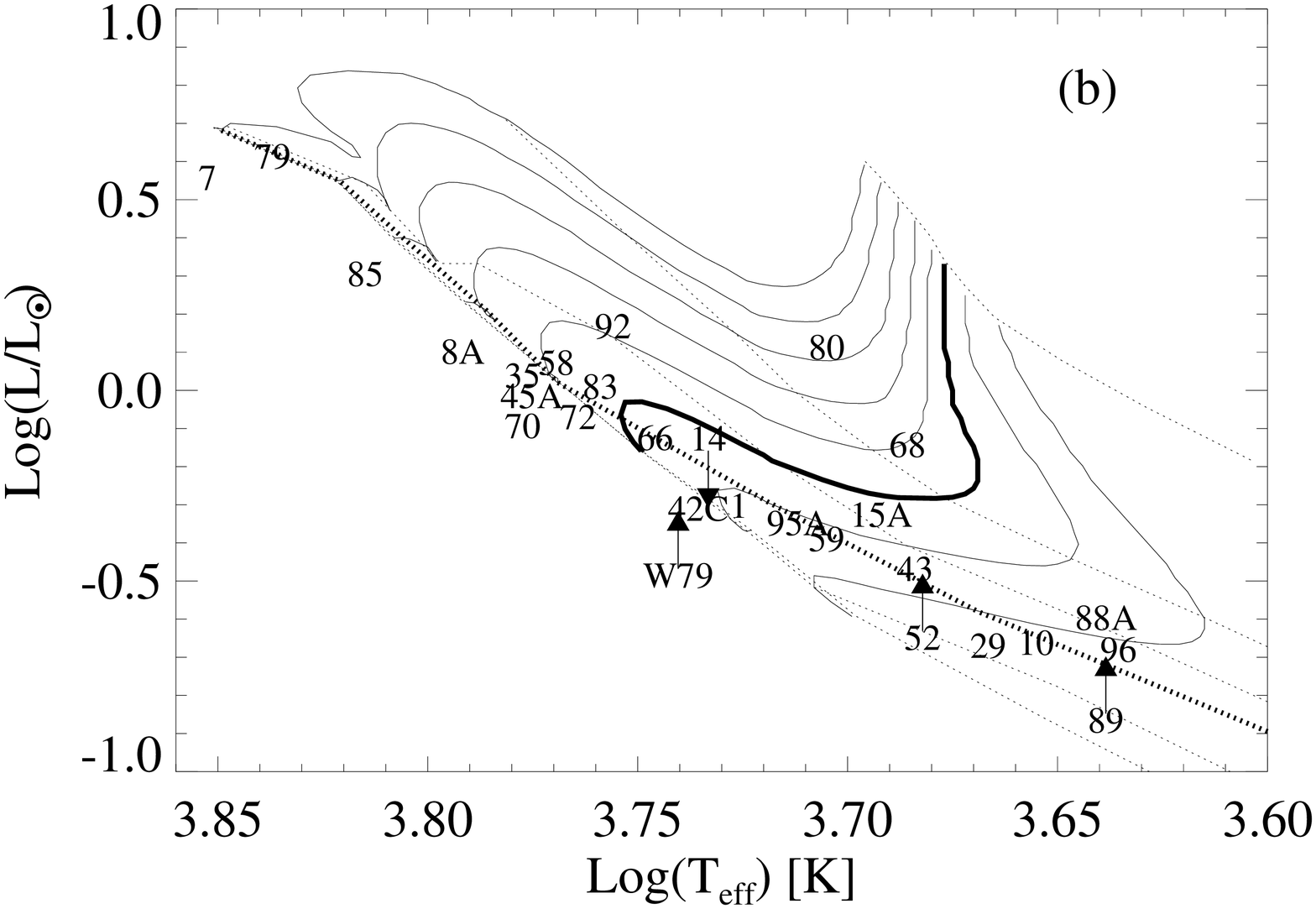}
	\caption{Evolutionary plots for the stars in (a) IC 2391 and (b) IC 2602. Only those stars which we determined as single and either classified as a member or a probable member (i.e.\ M or ? in Tables~\ref{irt2391rot} and~\ref{irt2602rot}) have been included. The VXR and R has been removed from the star name. Evolutionary tracks and Isochrones of \citet{DAntonaF:1997} are shown. The solid lines are evolutionary tracks for 0.8 \Msolar \ to 1.5 \Msolar \ (with the thick line representing 1.0 \Msolar). The dashed lines are age isochrones for 1, 10, 20, 30, 50, and 100 Myrs (with the thick dashed line representing 30 Myrs).}
	\label{irtHR}
	\end{center}
\end{figure}

Masses of the observed stars were estimated from \citet{DAntonaF:1997} isochrones to the nearest 0.1 solar mass (see Figure~\ref{irtHR}). Radii were then calculated from Equation~\ref{irtmbol} and: 
\BE
{\rm M_{bol} = 4.74 - 2.5 log(\frac{T_{eff}^{4} R^{2}}{T_{eff\odot}^{4} R_{\odot}^{2}})}
\EE

\subsection{Measuring Chromospheric Emission \label{irtmeasuringemission}}

Only those stars deemed to be probable cluster members and probably single (or at least single-lined spectroscopic binaries) had their chromospheric emission measured. Single stars were used because of the difficulty in determining the relative contributions to the emission for binary components. Before any chromospheric analysis was carried out on the spectra, telluric lines in the wings of the three IRT lines were first removed from both the target and inactive star spectra.

As mentioned previously there have been several studies involving the use of the IRT lines as chromospheric diagnostics, however the studies have adopted slightly different methods of determining the chromospheric activity of the stars, employing various bandwidths to measure chromospheric emission. The method we have adopted is that used by \citet{SoderblomDR:1993} in their measurement of the 8542\AA\/ line of solar-type stars in the Pleiades, whereby the chromospheric emission is calculated from the emission profile of the star.

\begin{figure}
	\begin{center}
	\includegraphics[angle=90,width=8.4cm]{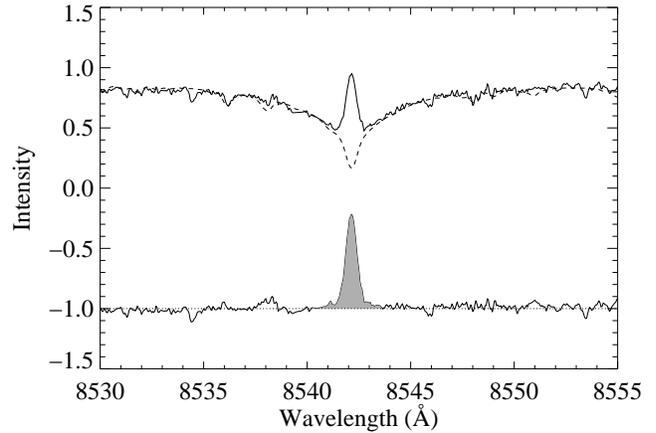}
	\caption{The emission profile for the 8542\AA\/ line of R89 (IC 2602). The dashed line is the fitted inactive star ($\alpha$ Cen B in this case) rotationally broadened to match the target star. The emission profile (Target - Inactive) is shown shifted by -1.0 for display purposes. The shaded area is the measured emission equivalent width.} 
	\label{irtdiff}
	\end{center}
\end{figure}

\begin{figure}
	\begin{center}
	\includegraphics[angle=90,width=8.4cm]{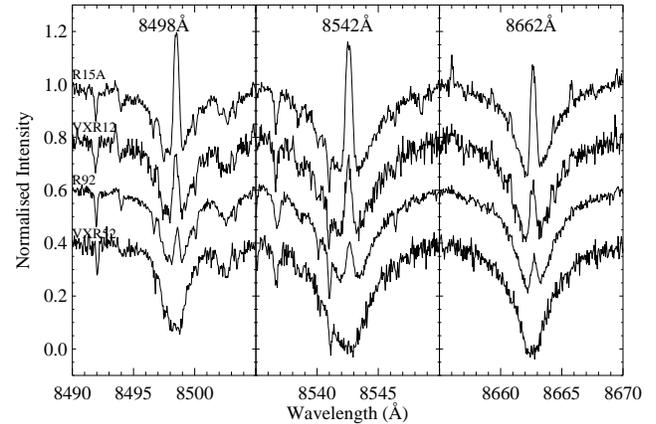}
	\caption{Examples of the Calcium infrared triplet lines for stars of decreasing activity. The stars were chosen to have similar \vsinis values and the profiles have been shifted down in 0.2 steps for display purposes.}
	\label{irtspectra}
	\end{center}
\end{figure}

For all three of the IRT lines the spectrum of an inactive star of similar spectral type was rotationally broadened to match that of the target star. The inactive star's spectrum was then shifted to match the radial velocity of the target star and was fitted to the target star's spectrum by matching the wings of the IRT line being examined, as the chromospheric emission should be limited to the core of the lines. The spectrum of the inactive star was then subtracted from the target star to produce an emission profile. To measure the chromospheric emission from the IRT lines, the area under the emission profile was calculated to produce an emission equivalent width. The process is outlined in Figure~\ref{irtdiff} for the 8542\AA\/ line while Figure~\ref{irtspectra} gives examples of the emission in the IRT lines for a sample of stars of differing activity levels.

By subtracting the IRT line of an inactive star (taken with the same instrumental setup) from that of our target stars, the photospheric contribution to the line is removed, and the resultant emission profile contains the chromospheric emission of the star. This method also removes a small amount of chromospheric emission, as no star is completely inactive. The inactive stars chosen for this project were
the same as those used to determine the \vsinis of the target stars, namely HD16673 (F6V), $\alpha$ Cen A (G2V), and $\alpha$ Cen B (K1V). These stars have near solar-metallicity (HD 16773: [Fe/H] = -0.11, $\alpha$ Cen A/B: [Fe/H] = 0.12 according to \citet{NordstromB:2004}). Although these inactive stars are not zero-age main-sequence stars (it is virtually impossible to get an inactive zero-age main-sequence star for comparison) the differences in surface gravity between these stars and those of IC 2391/2602 should have minimal effect on the chromospheric measurements, particularly for rapidly-rotating stars where rotational broadening will dominate the line profile.  

To measure the chromospheric emission of the IRT lines we have determined the chromospheric emission ratio (R$^{\prime}$\subs{IRT}) for each of the IRT lines. This gives the fraction of each star's bolometric luminosity that is emitted from the chromosphere in the IRT line. This is analogous to the chromospheric emission ratio R$^{\prime}$\subs{HK} derived from Ca II H and K observations described by \citet{NoyesRW:1984}, where the prime signifies that the photospheric emission has been removed. R$^{\prime}$\subs{8542}, the chromospheric emission ratio in the 8542\AA\/ line, has been used previously by \citet{SoderblomDR:1993} and \citet{JamesDJ:1997}.

To calculate R$^{\prime}$\subs{IRT} for each of the lines, the emission equivalent widths were first converted into the surface flux emitted by the IRT lines (F$^{\prime}$\subs{IRT}). Using the effective temperatures of the target stars from Tables~\ref{irt2391rot} and~\ref{irt2602rot}, the surface flux per \AA ngstrom was calculated for each star by interpolating between model atmospheres from  \citet{KuruczRL:1993}. This was then multiplied by the star's Emission Equivalent Width to give the surface flux in each of the IRT lines (F$^{\prime}$\subs{IRT}). In order to remove the dependence of the surface flux upon the star's colour, the chromospheric emission ratio in the IRT lines (R$^{\prime}$\subs{IRT}) was then calculated for each star. This is defined as the ratio of the surface flux in the IRT line to the star's total bolometric emission.
\BE
{\rm R^{\prime}_{IRT} = F^{\prime}_{IRT} / \sigma T^{4}_{eff}},
\EE
with F\sups{\prime}\subs{IRT} in W m\sups{-2} and T\subs{eff} in K. 

The chromospheric emission ratios, emission equivalent widths, and surface fluxes, for single members of the two clusters are given in Tables~\ref{irt2391chromo} and~\ref{irt2602chromo}.

\begin{table*}
\caption{}
\vbox to 220mm{\vfil Landscape Table 4 to go here.
\vfil}
\label{irt2391chromo}
\end{table*}

\begin{table*}
\caption{}
\vbox to 220mm{\vfil Landscape Table 5 to go here.
\vfil}
\label{irt2602chromo}
\end{table*}

\subsubsection{Errors in Chromospheric Emission Measures \label{irterrors}}

\begin{figure}
	\begin{center}
	\includegraphics[width=8.4cm]{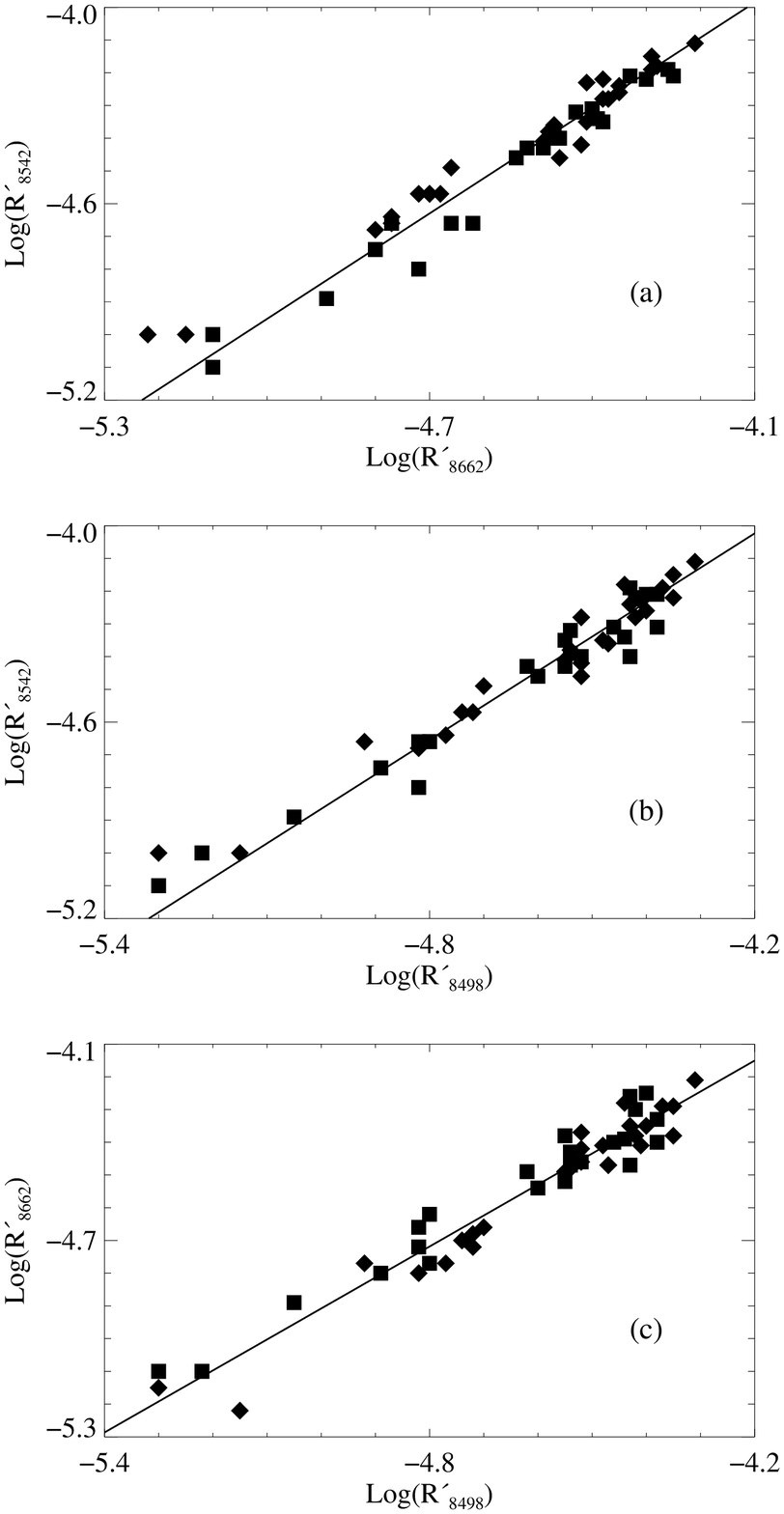}
	\caption{Comparison of the chromospheric emission ratios for the three infrared triplet lines, (a) Log(R\sups{\prime}\subs{8542}) vs Log(R\sups{\prime}\subs{8662}), (b) Log(R\sups{\prime}\subs{8542}) vs Log(R\sups{\prime}\subs{8498}), and (c) Log(R\sups{\prime}\subs{8662}) vs Log(R\sups{\prime}\subs{8498}). As in Figure~\ref{irtvsinicomp}, squares represent stars from IC 2391 and diamonds IC 2602. The lines are simple linear fits to the data and are given in Equations~\ref{irt_8542_8662},~\ref{irt_8542_8498}, and~\ref{irt_8662_8498}.}
	\label{irt_8662_8542_8498}
	\end{center}
\end{figure}

Possible errors in the measurement of the chromospheric emission come from a number of sources; errors in the measurement of the emission equivalent width, errors in determination of effective temperature, a mismatching of the spectral-type of the inactive stars to that of the target stars, and lastly the intrinsic variability of the chromospheric emission of the star's themselves. Errors in effective temperature determinations of $\pm$100 K are likely to produce errors of less than 10\% in the measurement of the chromospheric emission ratios, while \citet{SoderblomDR:1993} found no perceptible differences in the infrared triplet lines of the inactive field stars they studied which cover a similar colour range to our data, thus a mismatch of the spectral-type of the inactive template star should have minimal effect of the results. Thus the most likely sources of error to affect our results are errors in the measuring of the emission equivalent width and the intrinsic variability of the stars themselves. Although Tables~\ref{irt2391rot} and~\ref{irt2602rot} show that multiple observations were taken of each star these were mostly taken consecutively, so unfortunately cannot be used to determine any measure of the intrinsic variability of the stars. In order to determine the error in the measurement of the emission equivalent width on the chromospheric emission ratios of these stars we have plotted a comparison of the chromospheric emission ratios for the three infrared triplet lines (see Figure~\ref{irt_8662_8542_8498}).  We have then fitted simple linear fits to the data and determined the following relationships:
\BE
{\rm Log(R^{\prime}_{8542}) = 1.07 \times Log(R^{\prime}_{8662}) + 0.42}
\label{irt_8542_8662}
\EE
\BE
{\rm Log(R^{\prime}_{8542}) = 1.05 \times Log(R^{\prime}_{8498}) + 0.40}
\label{irt_8542_8498}
\EE
\BE
{\rm Log(R^{\prime}_{8662}) = 0.95 \times Log(R^{\prime}_{8498}) - 0.17}
\label{irt_8662_8498}
\EE

Looking at the scatter around these lines of best fit it can be seen that the error in Log(R\sups{\prime}\subs{8542}) when plotted against Log(R\sups{\prime}\subs{8662}), i.e.\ comparing the two strongest lines, is smaller for more active stars than for inactive stars and varies from $<$$\pm$0.1 for active stars up to $\sim\pm$0.2 for the inactive stars in the sample. This error is actually a cumulative measure of the error in both of the lines plotted and thus the measurement error in each individual line is most likely less than this. However, taking into account possible errors in the effective temperature we believe that $\pm$0.1 is a reasonable estimate of the error in our Log(R\sups{\prime}\subs{IRT}) values. An error of $\pm$0.1 in Log(R\sups{\prime}\subs{IRT}) appears to hold even for the ultra-rapid rotators in the clusters, however the fitting of the inactive star spectrum is difficult due to the rotationally broadened width of the chromospheric lines in these stars, and thus based on fitting errors we believe that the error in Log(R\sups{\prime}\subs{IRT}) for the ultra-rapid rotators may be slightly higher than this at $\pm$0.15.

\subsection{Activity versus Rotation \label{irtactivityrotation}}

\begin{figure}
	\begin{center}
	\includegraphics[angle=90,width=8.4cm]{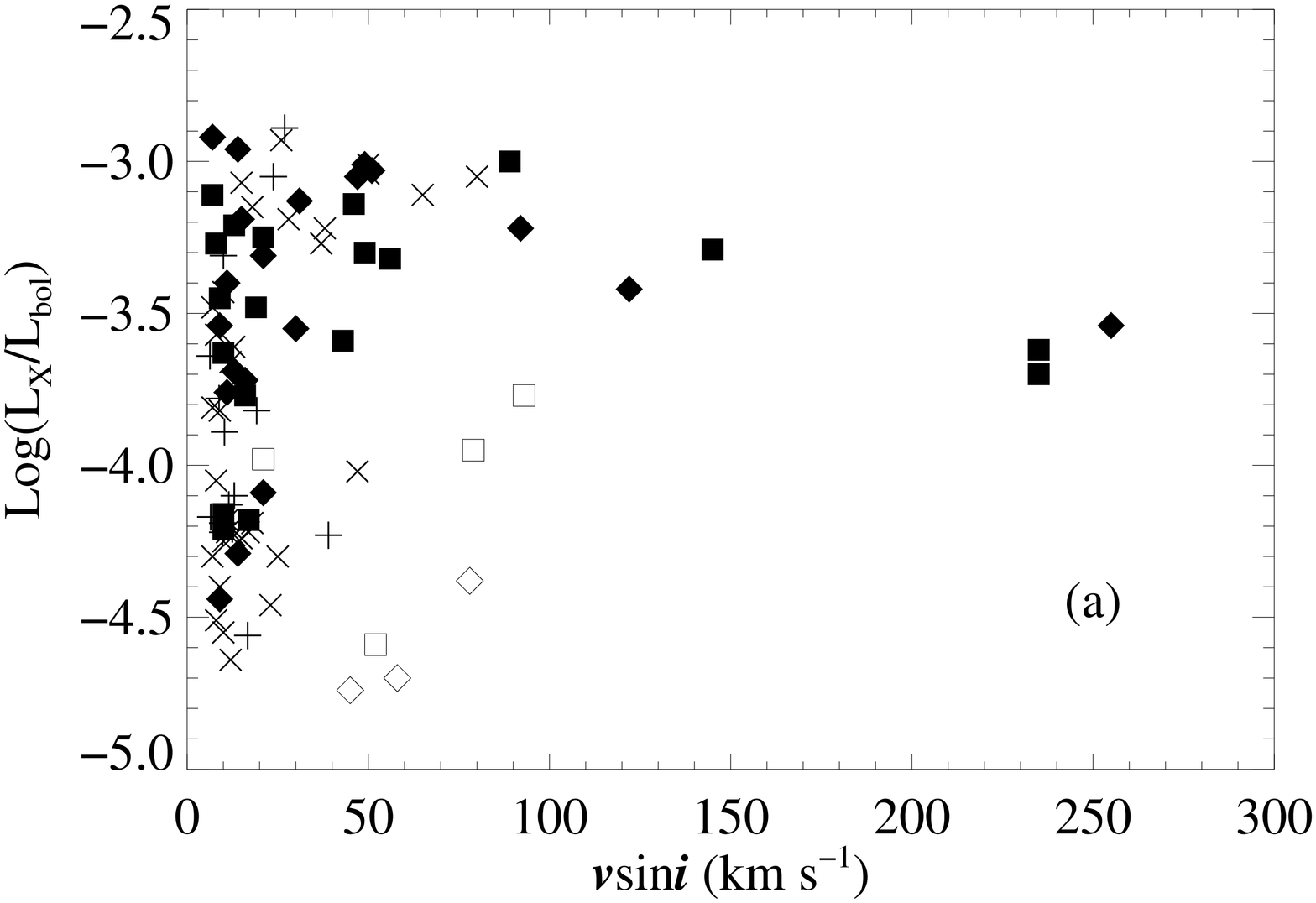}
	\includegraphics[angle=90,width=8.4cm]{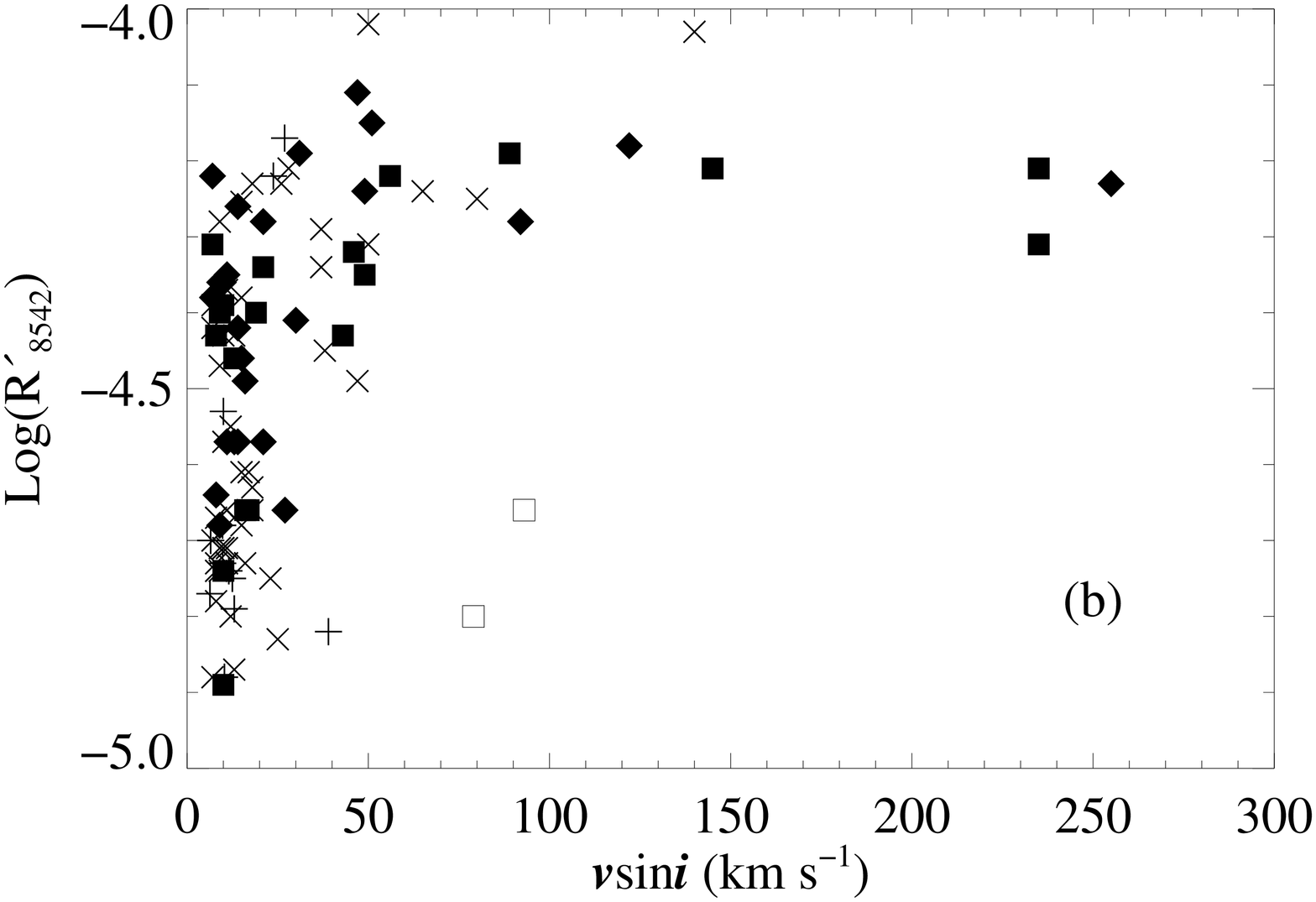}
	\caption{(a) The Log of the coronal X-ray emission divided by the star's bolometric luminosity and (b) the Log of the chromospheric emission ratio in the 8542\AA\/ line, both versus \vsinis for the observed single members of IC 2391 and IC 2602. Again squares represent stars from IC 2391 and diamonds IC 2602. Filled and open symbols represent stars with (V-I\subs{C})\subs{0} $>$ 0.6 and (V-I\subs{C})\subs{0} $<$ 0.6 respectively. Included in the plot are results from the Pleiades (crosses) and NGC 6475 (pluses). For (b) all stars with Log(R\sups{\prime}\subs{8542}) $\le$ -5.0 are not plotted (all such stars have (V-I\subs{C})\subs{0} $<$ 0.6). } 
	\label{irt8542vsini}
	\end{center}
\end{figure}

As mentioned previously the coronal X-ray emission of young active solar-type stars show a saturation of activity for rapid rotation, with the emission showing a further supersaturation effect for ultra-rapid rotation. Figure~\ref{irt8542vsini} plots the Log of the coronal X-ray emission (divided by the star's bolometric luminosity) and the Log of the chromospheric emission ratio for the strongest of the IRT lines, 8542\AA, against \vsinis for the observed single stars in IC 2391 and IC 2602.  The coronal X-ray data for IC 2391 and IC 2602 are taken from \citet{PattenBM:1996} and \citet{StaufferJR:1997} and were not taken simultaneously with the choromspheric results presented here. This means that variability and flaring may influence the scatter in these plots (and those in Figures~\ref{irt8542ross} and~\ref{irt8542Xray}). The \vsinis values used in Figure~\ref{irt8542vsini} are those measured in this study. The coronal data for the Pleiades were taken from \citet{StaufferJR:1994} and \citet{MicelaG:1999} while the chromospheric data were taken from \citet{SoderblomDR:1993}. For NGC 6475 both the coronal and chromospheric data were taken from \citet{JamesDJ:1997}.

Note: For the three stars in Tables~\ref{irt2391rot} and~\ref{irt2602rot} that have values of \vsinis $\le$ 7 \kmss and are plotted in this section (VXR76A, R1, and R15A), their \vsinis values have been assumed to be 7 \kmss for plotting purposes and the calculation of Rossby number if the rotational period is unknown.

\begin{figure}
	\begin{center}
	\includegraphics[angle=90,width=8.4cm]{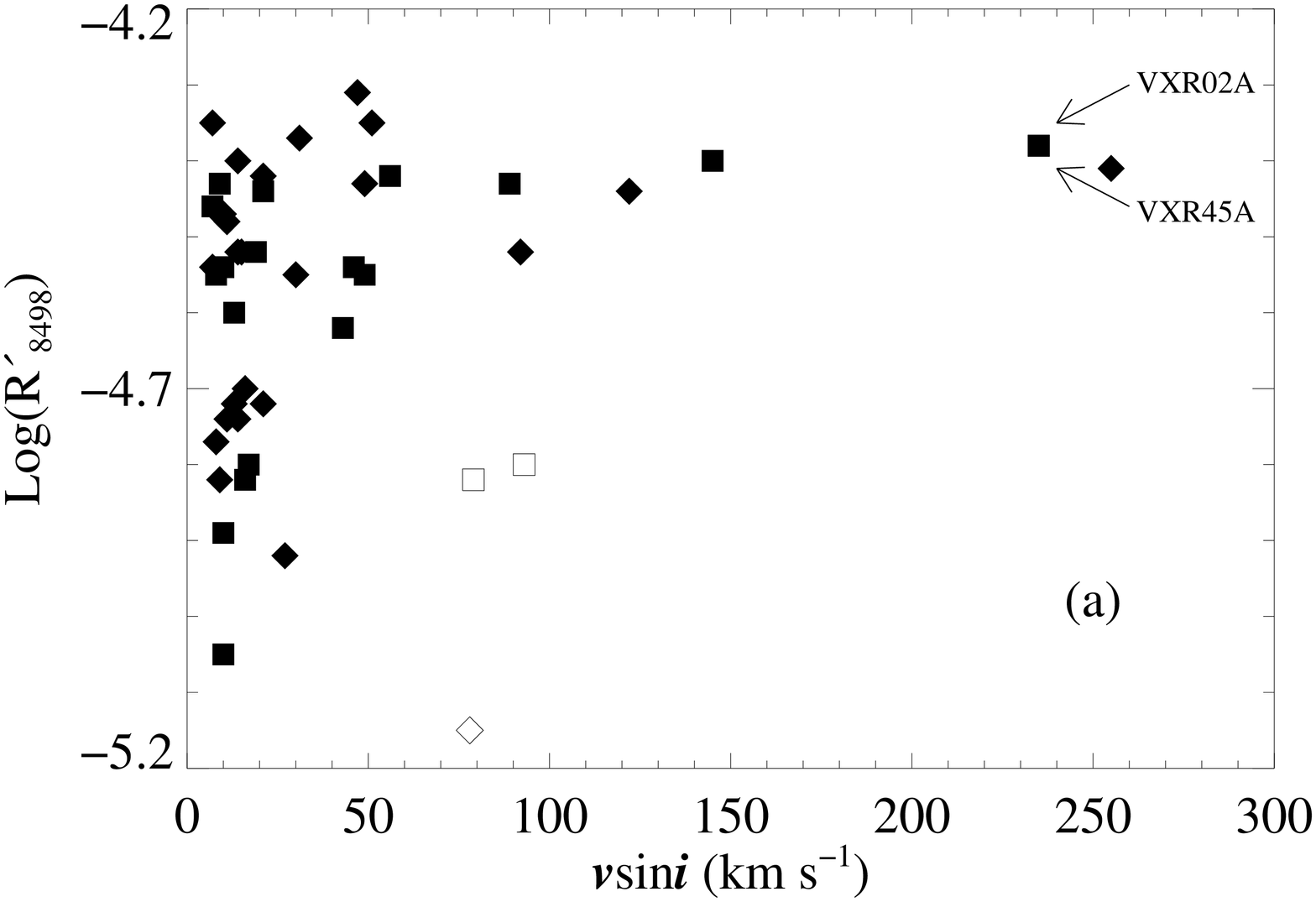}
	\includegraphics[angle=90,width=8.4cm]{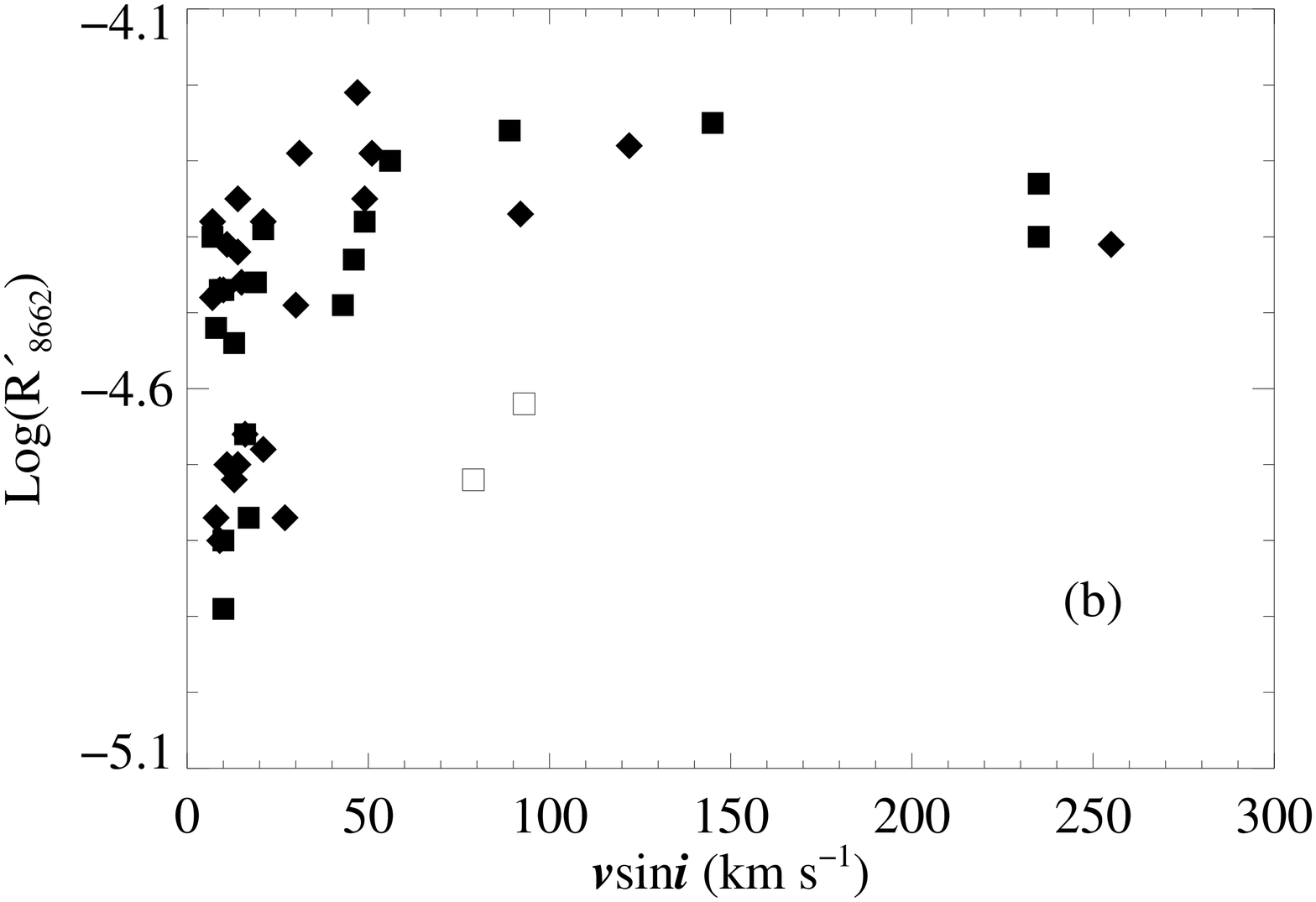}
	\caption{The chromospheric emission ratio in (a) the 8498\AA\/ line and (b) the 8662\AA\/ line versus \vsinis for our IC 2391/2602 stars. As with Figure~\ref{irt8542vsini} all stars with log(R\sups{\prime}\subs{8498}) and log(R\sups{\prime}\subs{8662}) $\le$ -5.2 and -5.1 respectively, are not shown. Again all these stars have (V-I\subs{C})\subs{0} $<$ 0.6. The symbols are the same as in Figure~\ref{irt8542vsini}.} 
	\label{irtothervsini}
	\end{center}
\end{figure}

\begin{figure}
	\begin{center}
	\includegraphics[angle=90,width=8.4cm]{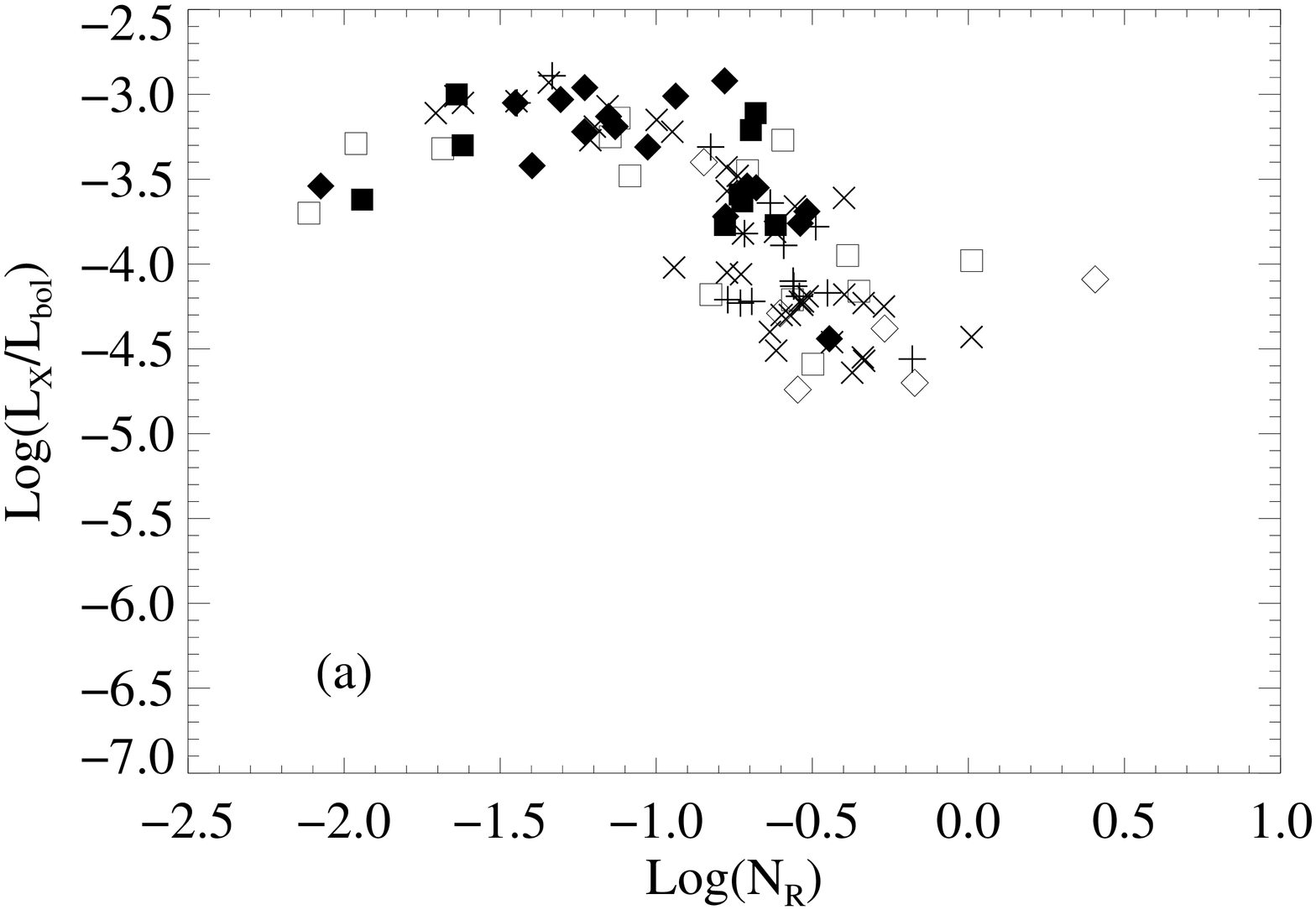}
	\includegraphics[angle=90,width=8.4cm]{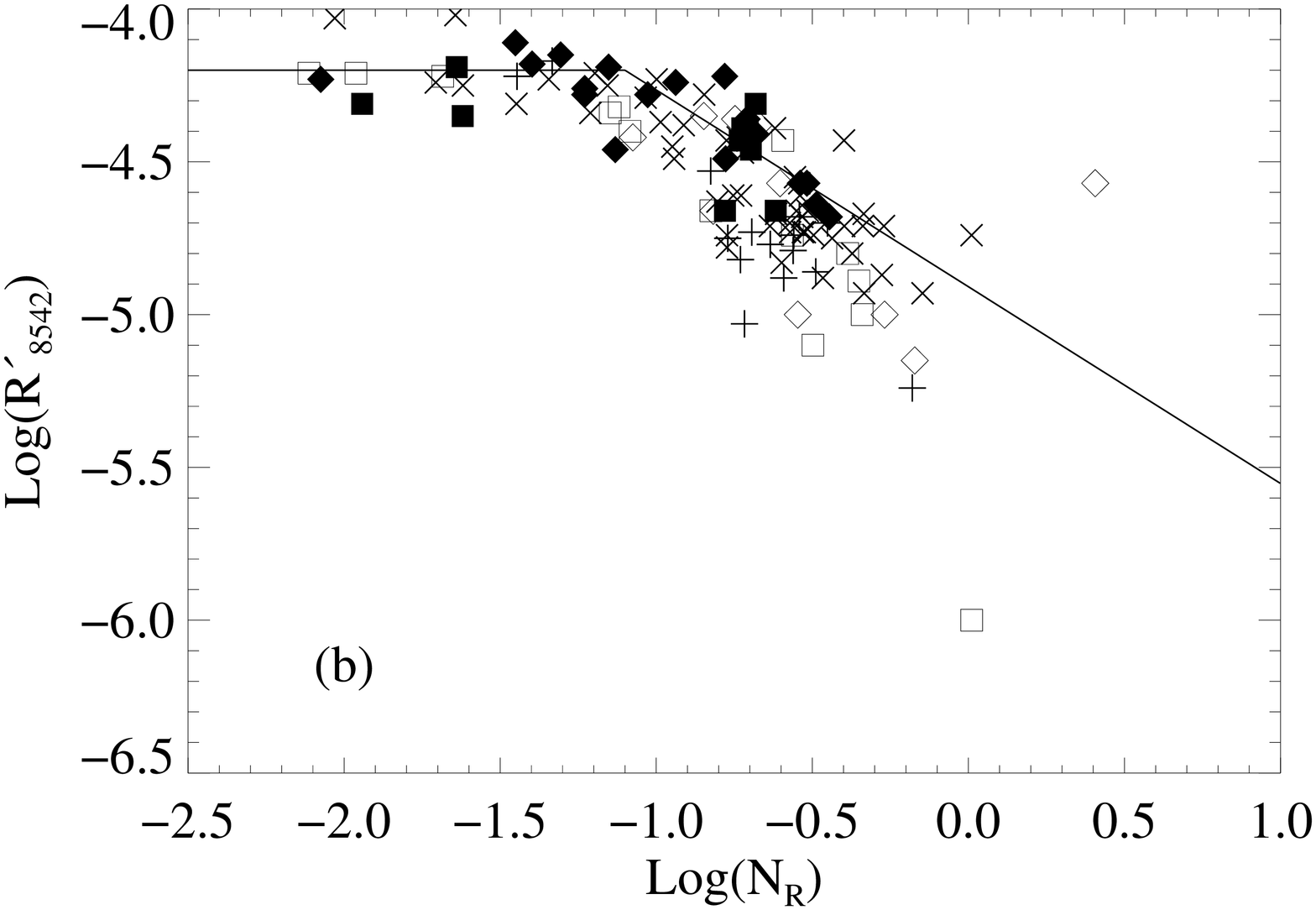}
	\caption{(a) The Log of the coronal X-ray emission divided by the star's bolometric luminosity, and (b) the Log of the chromospheric emission ratio in the 8542\AA\/ line, versus the Log of the Rossby number (N\subs{R}) for the single observed members of IC 2391 and IC 2602. Again results from the Pleiades (crosses) and NGC 6475 (pluses) are included and squares represent stars from IC 2391 and diamonds IC 2602. Filled and open symbols represent stars with known periods and stars whose Rossby number had to be estimated from Equation~\ref{irtcalross}, respectively. The black line is just an estimate of the saturation level for Log(N\subs{R}) $<$ -1.1, while for Log(N\subs{R}) $>$ -1.1 the line is a simple linear fit to the IC 2391/2602 data points with a fixed endpoint of Log(N\subs{R}) $=$ -1.1 and Log(R$^{\prime}$\subs{8542}) $=$ -4.2}.
	\label{irt8542ross}
	\end{center}
\end{figure}

The activity levels of the target stars have also been determined for the two weaker IRT lines (8498\AA\/ and 8662\AA). The Log of the chromospheric emission ratio versus \vsinis for these two lines is given in Figure~\ref{irtothervsini}.

Both the chromospheric (in all three of the IRT lines) and the coronal emission show a saturation for rapid rotators. The coronal emission shows supersaturation for ultra-rapid rotators. However, the chromospheric emission shows little (if any) decrease for the ultra-rapid rotators (although the 8662\AA\/ line appears to show a possible slight decrease).

In order to remove both the uncertainty in the inclination angle as well as any mass dependency in the velocity at which saturation starts (the saturation velocity),  \citet{NoyesRW:1984} plotted activity against a value called the Rossby number (N\subs{R}) as a measure of a star's rotation rate. Rossby number is defined as the star's rotational period divided by a measure of the star's convective turnover timescale ($\tau$\subs{c}). \citet{NoyesRW:1984} determined $\tau$\subs{c} based on the (B-V) colour of the target star. 

A number of the stars in Tables~\ref{irt2391rot} and~\ref{irt2602rot} have period determinations. The Rossby number for these stars has been calculated using $\tau$\subs{c} from \cite{NoyesRW:1984} Equation 4:
\BE
{\rm log (\tau_{c}) = \left\{\begin{array}{ll}
			1.362 - 0.166x + 0.025x^{2} - 5.323x^{3} & x > 0 \\
			1.362 - 0.14x                            & x < 0
			\end{array}
			\right. }\label{irttauc}
\EE
where $x$ = 1 $-$ (B-V)\subs{0} and $\tau$\subs{c} is in days. There have been later calculations of the convective turnover timescale, such as those by \citet{KimYC:1996} used by \citet{KrishnamurthiA:1998}, which are more suited to pre-main sequence stars. The \citet{KimYC:1996} values for $\tau$\subs{c} do vary from those calculated using the \citet{NoyesRW:1984} formula above. If the \citet{KimYC:1996} values are used, Figure~\ref{irt8542ross} remains fundamentally unchanged except that the Rossby numbers are all shifted to lower values and the saturation occurs at Log(N\subs{R}) $\sim$ -1.5 rather than the Log(N\subs{R}) $\sim$ -1.0 shown in Figure~\ref{irt8542ross}.  We have decided to use the \citet{NoyesRW:1984} values as many of the previous observations of saturation and supersaturation have used them \citep[e.g.][]{SoderblomDR:1993,PattenBM:1996,StaufferJR:1997}.

For those stars in Tables~\ref{irt2391rot} and~\ref{irt2602rot} that do not have measured rotational periods, we have used the method of \citet{SoderblomDR:1993} to estimate the Rossby number from the
\vsinis of the star. The equation used is:
\BE
{\rm N_{R} = \frac{2 \pi R}{\tau_{c} vsini}} \label{irtcalross}
\EE
where R is the stellar radius (in km), \vsinis is in \kmsn, and $\tau$\subs{c} (in seconds) is calculated from Equation~\ref{irttauc}. Assuming that stellar inclination angles are randomly distributed, then $<$sin$i$$>$ should be 0.785, so we have multiplied Equation~\ref{irtcalross} by this amount. We have also done this for the calculated Rossby numbers for the Pleiades and NGC 6475 which we have plotted.

\begin{figure}
	\begin{center}
	\includegraphics[width=8.4cm]{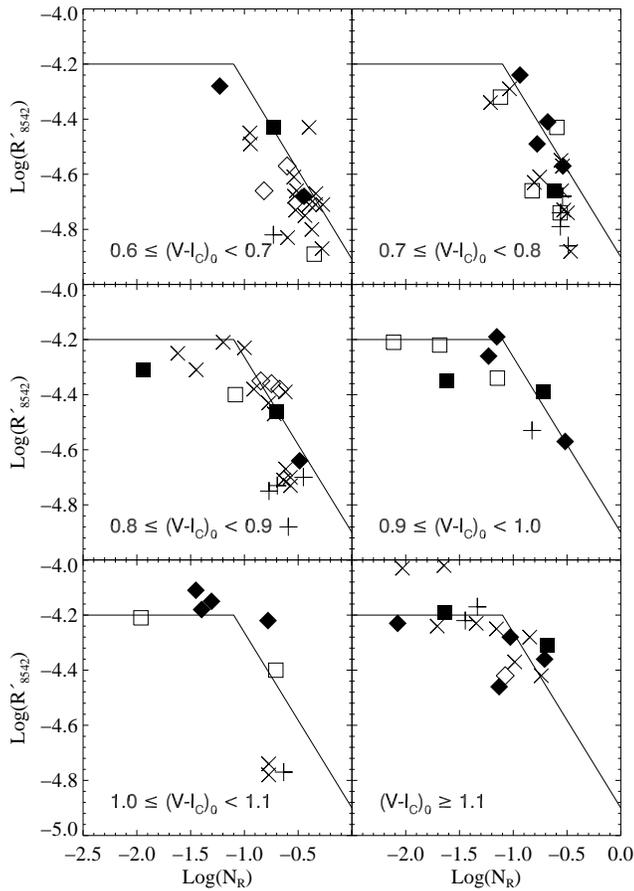}
	\caption{The Log of the chromospheric emission ratio in the 8542\AA\/ line, versus the Log of the Rossby number (N\subs{R}) for the single observed members of IC 2391 and IC 2602, separated into colour bins.  Only those stars with  (V-I\subs{C})\subs{0} $\ge$ 0.6 have been included. The symbols are the same as in Figure~\ref{irt8542ross}, however the scale of the plots is different.}
	\label{irt8542rosscol}
	\end{center}
\end{figure}

\begin{figure}
	\begin{center}
	\includegraphics[angle=90,width=8.4cm]{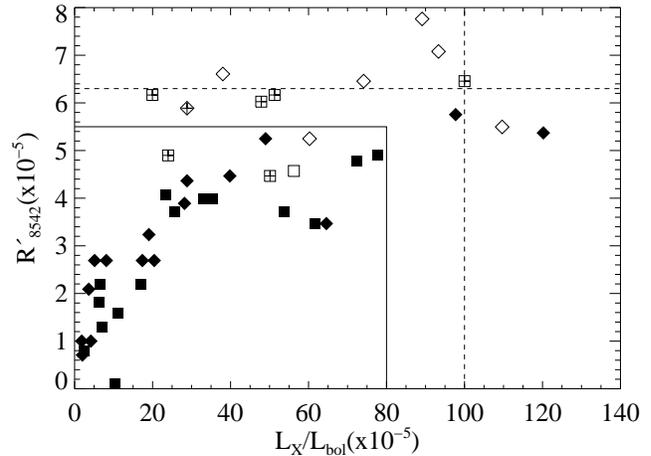}
	\caption{Chromospheric emission ratio in the 8542\AA\/ line versus coronal X-ray activity. Again squares represent IC 2391 stars and diamonds IC 2602 stars. The dashed lines show the approximate saturation levels for the activity indicators. Open symbols represent stars with Log(N\subs{R}) $\le$ -1.14 and those symbols with a plus sign inside represent those stars with Log(N\subs{R}) $\le$ -1.62.}
	\label{irt8542Xray}
	\end{center}
\end{figure}

Figure~\ref{irt8542ross} plots the Log of the coronal X-ray emission (divided by the star's bolometric luminosity) and the Log of the chromospheric emission ratio for 8542\AA\/ line against the Log of the Rossby number for our stars in IC 2391/2602 with Figure~\ref{irt8542rosscol} showing how the chromospheric emission changes with colour (unsurprisingly showing that redder stars are more active). Again, in both Figures, the coronal X-ray data for IC2391/2602 are from \citet{PattenBM:1996} and \citet{StaufferJR:1997}, the coronal and chromospheric data for the Pleiades are from \citet{StaufferJR:1994}, \citet{MicelaG:1999}, and \citet{SoderblomDR:1993}, and the NGC 6475 data are from \citet{JamesDJ:1997}.

As was shown in Figures~\ref{irt8542vsini} and~\ref{irtothervsini}, both the coronal and the chromospheric emission show a saturation level. Again, the coronal emission shows evidence of supersaturation (with a decrease in activity for Log(N\subs{R}) $\la$ -1.7), whereas the chromospheric emission shows little evidence of a decrease. The chromospheric saturation level is around Log(R$^{\prime}$\subs{8542}) $\sim$ -4.2. This is similar to that shown in the Ca II 8542\AA\/ emission for the Pleiades stars by \citet{SoderblomDR:1993} and the NGC 6475 stars by \citet{JamesDJ:1997}. The scatter that the unknown inclination angle and estimated radius adds to Figure~\ref{irt8542ross} for stars without a  measured rotation period does not appear to be significant compared with the inherent scatter in the emission of the stars.

The onset of saturation appears to occur at around Log(N\subs{R}) $\sim$ -1.0 for both the coronal and chromospheric emission. If saturation occurs at the same Rossby number then the relationship between the coronal and chromospheric emission should be linear approaching saturation. Ignoring those stars with activity around the saturation level (in either coronal or chromospheric emission), i.e.\ including only those stars in the box in Figure~\ref{irt8542Xray}, there appears to be a change in the relationship between the two activity indicators for the stars in this study, between stars with L\subs{X}/L\subs{bol} $\la$ 20 x 10\sups{-5} (Log(L\subs{X}/L\subs{bol}) $\la$ -3.7) and those stars above this value. This change occurs below both the coronal and chromospheric saturation level and could indicate that coronal and chromospheric emission have a slightly different relationship for stars with low activity than for more active stars, rather than indicating a change in the onset of saturation. For stars more active than L\subs{X}/L\subs{bol} $\sim$ 20 x 10\sups{-5} the relationship between the coronal and chromospheric emission (up to the saturation levels) could well be linear, but the scatter in the data is too large to determine if this is the case.

Ignoring the supersaturated stars, all but two of the stars within the box in Figure~\ref{irt8542Xray} (and two outside) have Log(N\subs{R}) $\ge$ -1.14. This indicates that the Rossby number for the onset of saturation (for both coronal and chromospheric emission) is lower than Log(N\subs{R}) = -1.0 indicated in Figure~\ref{irt8542ross} and is probably closer to Log(N\subs{R}) = -1.1. In addition, Figure~\ref{irt8542Xray} shows that all but three stars with both a small Rossby number,  Log(N\subs{R}) $\le$ -1.14, and low coronal activity, L\subs{X}/L\subs{bol} $\le$ 60 x 10\sups{-5}, (i.e.\ supersaturated) have Log(N\subs{R}) $\le$ -1.62, with only one star having a ``normal'' saturated level of coronal activity and a Log(N\subs{R}) $\le$ -1.62.  This appears to indicate that coronal supersaturation is occurring at a slightly higher Rossby number than shown in Figure~\ref{irt8542ross}(a), possibly closer to log(N\subs{R}) = -1.6.

\subsection{Colour and Age Effects \label{irtageeffects}}

\begin{figure}
	\begin{center}
	\includegraphics[angle=90,width=8.4cm]{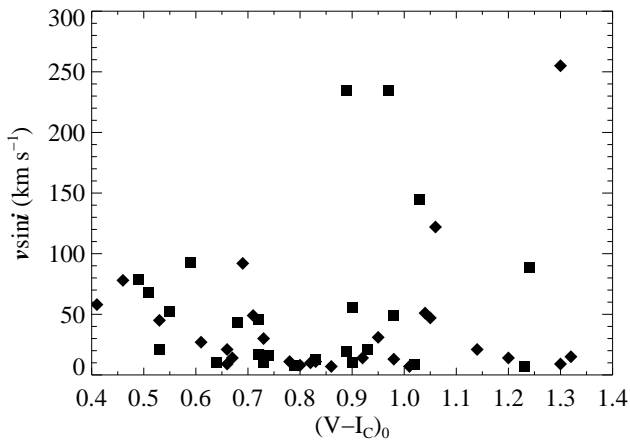}
	\caption{Measured \vsinis values versus colour. The symbols are the same as in Figure~\ref{irtvsinicomp}.}
	\label{irtvsinicol}
	\end{center}
\end{figure}

By the age of the Hyades $\sim$ 625 $\pm$ 50 Myrs \citep{PerrymanMAC:1998}, the rotation rate of a solar-type star can be reasonably determined based on its colour, with more massive stars showing higher rotation rates \citep{RadickRR:1987}. However, due to the minimal amount of magnetic braking they have undergone, solar-type stars of the age of IC 2391 and IC 2602 should show a wide range of rotation rates for all colours. Figure~\ref{irtvsinicol} plots \vsinis against colour for the single stars in IC 2391 and IC 2602. As expected there is a wide range of \vsinis values (up to \vsinis = 100 \kmsn) in the clusters regardless of the star's colour, however the ultra-rapid rotators in the clusters are all redder than (V-I\subs{C})\subs{0} $\sim$ 0.85. There is a small dip in the data around (V-I\subs{C})\subs{0} $\sim$ 0.8, showing that there is a paucity of rapid rotators observed for this colour. We suggest that this is most likely just a statistical effect.

\begin{figure}
	\begin{center}
	\includegraphics[width=8.4cm]{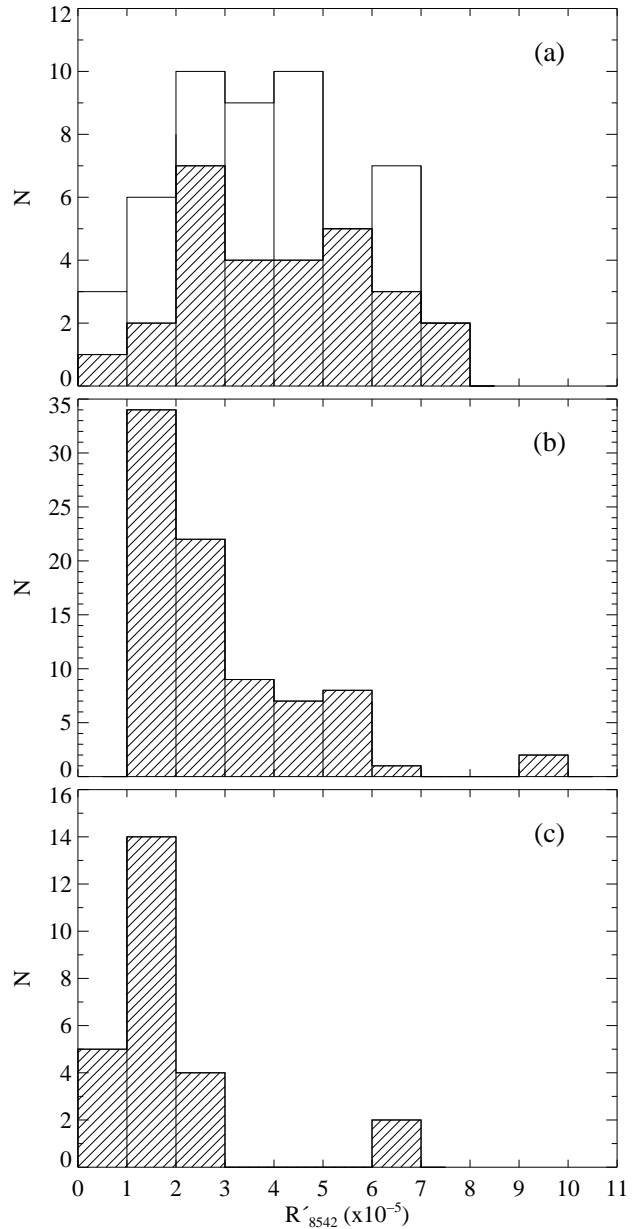}
	\caption{Histogram of the distribution of chromospheric emission ratios in the 8542\AA\/ line for (a) the single members of IC 2391 \& IC 2602, (b) the Pleiades, and (c) NGC 6475. The shaded data in (a) represents IC 2602 while the data with no shading represents IC 2391. Only stars with measured \vsinis have been included.}
	\label{irt8542hist}
	\end{center}
\end{figure}

As stars age, their activity level falls as their rotation rate decreases through the loss of angular momentum via magnetic braking, i.e.\ \citet*{BouvierJ:1997}. Therefore it is expected that the overall activity levels of clusters should progressively shift to lower values for clusters of increasing age. There are two other young clusters for which the chromospheric emission ratio in the 8542\AA\/ line has been measured, the Pleiades \citep{SoderblomDR:1993} at an age of 130 $\pm$ 20 Myrs according to the Lithium-depletion boundary method \citep{BarradoD:2004} and NGC 6475 \citep{JamesDJ:1997} at an age of $\sim$220 Myrs. Figure~\ref{irt8542hist} gives a histogram of the chromospheric emission ratios for IC 2391 and IC 2602, along with the ratios for the Pleiades and NGC 6475 clusters. The data for the Pleiades were taken from \citet{SoderblomDR:1993} while the data for NGC 6475 were taken from \citet{JamesDJ:1997}. Most stars in IC 2391 and IC 2602 are seen to have a chromospheric emission level of around R$^{\prime}$\subs{8542} $\sim$ 2 to 5 x 10\sups{-5} compared to the Pleiades and NGC 6475, in which the peak is around R$^{\prime}$\subs{8542} $\sim$ 1 to 2 x 10\sups{-5}. The median for IC 2391/2602 is R$^{\prime}$\subs{8542} $\sim$ 4.0 x 10\sups{-5}, for the Pleiades the median is R$^{\prime}$\subs{8542} $\sim$ 2.2 x 10\sups{-5} (although this may be skewed by the fact that no star has R$^{\prime}$\subs{8542} $<$ 1.0 x 10\sups{-5}) and the median is R$^{\prime}$\subs{8542} $\sim$ 1.7 x 10\sups{-5} for NGC 6475. As expected the median chromospheric activity level for the clusters drops with increasing cluster age, with IC 2391/2602 showing a much larger spread in chromospheric activity compared to the older clusters.

\begin{figure}
	\begin{center}
	\includegraphics[width=8.4cm]{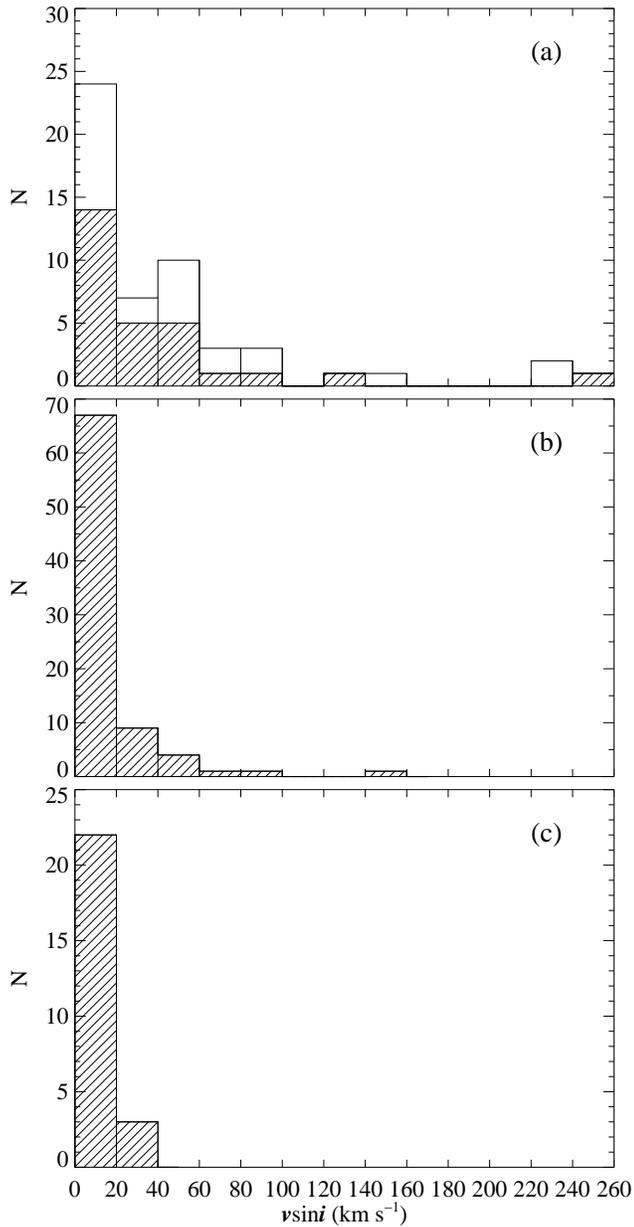}
	\caption{Histogram of the distribution of \vsinis for the single members observed in (a) IC 2391 \& IC 2602, (b) the Pleiades, and (c) NGC 6475. The shading and references are the same as in Figure~\ref{irt8542hist}. Only stars with R$^{\prime}$\subs{8542} measures have been included.}
	\label{irtvsinihist}
	\end{center}
\end{figure}

At the age of IC 2391 and IC 2602 the solar-type stars in the two clusters should have undergone no significant main-sequence braking and should represent the rotation rate with which they arrive (or will soon arrive) on the ZAMS. Figure~\ref{irtvsinihist} is a histogram of the measured \vsinis values for the single stars observed in the two clusters in this study, along with the \vsinis distributions for the Pleiades and NGC 6475. Again the Pleiades data are from \citet{SoderblomDR:1993} and the NGC 6475 data are from \citet{JamesDJ:1997}. The number of slow rotators compared to fast rotators, using the division of \vsinis = 20 \kmsn, is approximately even for IC 2391/2602 with 24 stars with \vsinis $<$ 20 \kmss and 28 stars with \vsinis $>$ 20 \kmsn. Thus $\sim$54\% are rapid rotators. This is much the same as the similarly aged cluster NGC 2547 (lithium depletion boundary age $=$ 35 Myrs, \citealt{JeffriesRD:2005}) which has 50\% - 60\% (depending on the inclusion of possible binaries or not) of its solar-type stars as rapid rotators \citep*{JeffriesRD:2000}. For the older clusters (Pleiades and NGC 6475) magnetic braking should have slowed many of the rapid rotators in the clusters. This can be seen in their \vsinis distributions, with the Pleiades having 67 stars with \vsinis $<$ 20 \kmss and only 14 stars with \vsinis $>$ 20 \kmsn, $\sim$18\% rapid rotators. For NGC 6475 there are 23 stars with \vsinis $<$ 20 \kmss and only 3 stars with \vsinis $>$ 20 \kmsn, $\sim$12\% rapid rotators. By the age of the Pleiades the median \vsinis of the cluster has decreased significantly from that of IC 2391/2602, with the median \vsinis of IC 2391/2602 = 21 \kmsn, the Pleiades = 9 \kmsn, and NGC 6475 = 8 \kmsn.

When looking at the difference between the G stars (0.6 $\le$ (B-V)\subs{0} $\le$ 0.9) and the early-K stars (0.9 $\le$ (B-V)\subs{0} $\le$ 1.2) in IC 2391/2602 the percentage of rapidly rotating G stars is 34\% while for early-K stars it is 58\%. For the Pleiades the percentages are 15\% for the G stars and 29\% for the early-K stars. In NGC 6475 the rapid rotators makeup 0\% of the G stars, but 33\% of the early-K stars. Thus between the ages of IC 2391/2602 and the Pleiades the proportion of rapidly rotating G and K stars drops by about half (with the proportion of rapidly rotating G stars dropping slightly more than for the early-K stars). By the age of NGC 6475 there are no rapidly rotating G stars while the proportion of rapidly rotating early-K stars is similar to that of the Pleiades (although this may be influenced by the small number of early-K stars observed in NCG 6475; 2 of the 6 observed are rapid rotators). 

The rotational evolution of cluster stars has been studied by a number of people including a recent study by \citet{IrwinJ:2007} which includes rotational periods for a number of young clusters of varying ages, including rotational data from IC 2391/2602 stars. The observations presented in \citet{IrwinJ:2007} show that the rotational period of G stars falls at a similar rate to that of K stars (from a peak around the IC 2391/2602 age) for clusters up to $\sim$200 Myrs. After this age the observations show that the rotation rate of G stars falls quite dramatically compared to that for K stars. This is in good agreement with the \vsinis results from IC 2391/2602, the Pleiades, and NGC 6475. 

\citet{IrwinJ:2007} have also modelled the angular momentum evolution of young clusters as they age, incorporating dynamo saturation into the models. The results indicate that a mass-dependent saturation velocity is required to reproduce the observations (taken into account via the mass-dependency of the Rossby number in our results, see \S~\ref{irtactivityrotation}). Additionally, \citet{IrwinJ:2007} found that while solid-body rotation can explain the evolution of the fastest rotators, core-envelope decoupling (with different rotations in the stellar core and envelope) appears to be required to explain the evolution of the slowest rotators.

\section{Discussion}\label{irtdiscussion}

\subsection{Chromospheric Saturation \label{irtsaturation}}

Figure~\ref{irt8542ross}(a) gives the relationship between the coronal X-ray emission of the single solar-type stars in IC 2391 and IC 2602 and Rossby number. Figure~\ref{irt8542ross}(b) gives the same plot for the chromospheric emission ratio in the 8542\AA\/ line. For stars with Log(N\subs{R}) $\la$ -1.1 there is a saturation in the chromospheric emission at a level of around Log(R$^{\prime}$\subs{8542}) $\sim$ -4.2. This agrees well with the results of \citet{SoderblomDR:1993} and \citet{JamesDJ:1997}, who found a saturation in the chromospheric emission ratio for solar-type stars in the Pleiades and NGC 6475 of a similar level. The saturation levels for the two other Calcium IRT lines at 8498\AA\/ and 8662\AA\/ are shown in Figure~\ref{irtothervsini} and are Log(R$^{\prime}$\subs{8498}) $\sim$ -4.4 and Log(R$^{\prime}$\subs{8662}) $\sim$ -4.3. This confirms that, as expected, there is a saturation in the chromospheric emission of these cluster stars similar to that observed in the X-ray data from this and other clusters.

If all activity indicators show the onset of saturation at the same value of Rossby number, this would support the observed saturation being caused by the same mechanism (i.e.\ dynamo saturation). Figure~\ref{irt8542ross} shows the onset of saturation is around Log(N\subs{R}) $\sim$ -1.0 to -1.1 for both the coronal X-ray emission and the chromospheric. The chromospheric emission in both H$\alpha$ and the 8542\AA\/ line from the Pleiades dwarfs also show the onset of saturation at around Log(N\subs{R}) $\sim$ -1.0 \citep{SoderblomDR:1993}. However, \citet{CardiniD:2007} show that for the Mg II emission in a range of stars (including those in IC 2602) the onset of saturation occurs around Log(N\subs{R}) $\sim$ -0.7 (with a saturation level of Log(F\subs{MgII}) $\sim$ 6.3). Thus the onset of saturation for the Mg II emission appears to occur at a slightly higher Rossby number than what we find for the Ca II emission but may be within the error bars as \citet{CardiniD:2007} calculated the Rossby number empirically themselves rather than using the \citet{NoyesRW:1984} value. In other work, the measurement of stellar activity using $\Delta$V (optical light curve amplitude, used as a proxy for covering fraction of starspots) by \citet{ODellMA:1995}, shows the onset of saturation only for Log(N\subs{R}) $\la$ -1.5, if at all (using the \citet{NoyesRW:1984} $\tau$\subs{c} values). This is long after both coronal and chromospheric saturation have set in, but more recent work by \citet{KrishnamurthiA:1998} has questioned the use of $\Delta$V as an activity indicator. 

Thus we conclude that both the chromospheric and coronal emission appear to be saturating at a similar Rossby number. This supports the idea that both coronal and chromospheric saturation are caused by the same mechanism.

What are the possible mechanisms behind saturation? \citet{JardineM:1999} have suggested that the effect of saturation seen in X-ray emission could be the result of a reduction in the coronal X-ray emitting volume due to centrifugal stripping of the corona for rapid rotators. As rotation rate increases, the co-rotation radius of the star decreases. \citet{JardineM:1999} suggest that around the saturation velocity the co-rotation radius moves inside the X-ray emitting corona. The rise in gas pressure then breaks the magnetic loops to form open field regions that are dark in X-rays. As rotation increases the reduction in emitting volume is balanced by a rise in the density of the corona leading to saturation. While this idea explains coronal (X-ray) saturation, it cannot be used to explain chromospheric saturation. To explain chromospheric saturation, \citet{JardineM:1999} suggest that another mechanism, such as the enhanced continuum emission proposed by \citet{DoyleJG:1996} may be responsible. However, if coronal and chromospheric saturation are controlled by two different mechanisms it would be unlikely (but not impossible) that they would saturate at the same Rossby number. Thus the \citet{JardineM:1999} explanation would appear to be unlikely.

Another suggestion to explain coronal saturation has been put forward by \citet{StepienK:2001} who suggest that coronal saturation is the result of the coronal emitting regions of a star being completely filled. This, we assume, would similarly be happening to the chromospheric emission regions to account for chromospheric saturation. In this scenario the onset of saturation could conceivably occur at similar Rossby number for both coronal and chromospheric emission, if both are caused by strong magnetic fields erupting through the stellar surface and extending into the chromosphere and corona. However, as explained below, this idea appears to fall down when trying to explain supersaturation. 

Finally it should be noted that a different explanation for saturation (but not supersaturation) has been put forward by \citet{BarnesS:2003}. He argues that saturation is just the effect of a different dynamo in operation in rapidly-rotating stars, a convective rather than an interface dynamo. Such an explanation is supported by the finding of large regions of azimuthal magnetic field on the surfaces of young rapidly-rotating solar-type stars \citep[i.e.][]{DonatiJF:2003,MarsdenSC:2006}, which have been explained by the operation of a dynamo distributed throughout the convection zone \citep{DonatiJF:2003}. \citet{BarnesS:2003} claims that saturated stars should all be located on what he calls the convective sequence \citep[see][Figure 2]{BarnesS:2003}. However, this is just saying that saturated stars are more rapidly-rotating than non-saturated stars (i.e.\ rapidly-rotating stars are more active) which is obvious and thus our results provide no way of determining if the \citet{BarnesS:2003} explanation is correct.  

\subsection{Chromospheric Supersaturation? \label{irtsupersaturation}}

A further complication to the problem of saturation is the effect of supersaturation. The X-ray emission from the IC 2391 and IC 2602 stars, when plotted against Rossby number in Figure~\ref{irt8542ross}(a), shows the supersaturation effect occurring around Log(N\subs{R}) $\sim$ -1.6. This has been seen in the X-ray emission of other young cluster stars  \citep{PizzolatoN:2003,StaufferJR:1997,PattenBM:1996,RandichS:1996,ProsserCF:1996b}, and has also been seen, to a lesser extent, in the X-ray emission of W UMa and M dwarf stars \citep{StepienK:2001,JamesDJ:2000}. This study included measurements of the chromospheric emission of 5 stars in IC 2391 and IC 2602 with (V-I\subs{C})\subs{0} $>$ 0.6 and \vsinis values $>$ 100 \kmsn, including 3 stars with \vsinis values $>$ 200 \kmsn. The chromospheric emission for the IC 2391 and IC 2602 stars, Figures~\ref{irt8542vsini}(b),~\ref{irtothervsini}, and~\ref{irt8542ross}(b), show little evidence of a decrease in chromospheric emission for stars with \vsinis $\ga$ 100 \kmss or Log(N\subs{R}) $\la$ -1.6. Thus, it would appear that either supersaturation does not exist for the chromospheric emission of ultra-rapid rotators, or it occurs at a rotation rate significantly greater than that for coronal emission. 

It should be noted that due to the massive broadening of the spectral lines with such high \vsinis values, the accuracy of fitting of the inactive star to the target star can have an impact on the level of emission measured. However, a change of Log(R$^{\prime}$\subs{8542}) from -4.2 to -4.5 (which would be required to have a similar supersaturation effect to that seen in the X-ray emission) is about twice what we estimate the error in Log(R$^{\prime}$\subs{8542}) to be (see \S~\ref{irterrors}) which we consider unlikely.

Our Calcium infrared triplet results are supported by the results of \citet{RandichS:1998} which show no supersaturation in the H$\alpha$ emission of ultra-rapid rotators in the 50-Myr old $\alpha$-Persei cluster. In addition, the Mg II study of cluster and field stars by \citet{CardiniD:2007} also shows no evidence of supersaturation, however the lowest Log(N\subs{R}) value they have for their stars is greater than -1.6, so they are not really within the supersaturation regime.

In contrast to our results, a study of a number of active stars of spectral type K2 or later by \citet{ScholzA:2007} has shown tentative evidence for supersaturation in the H$\alpha$ emission of two stars with \vsinis values around 100 \kmsn. However, these are true pre-main sequence objects and the swollen convection zone of such objects may act differently to that of more zero-age main-sequence stars we have studied. This is supported by \citet{FeigelsonED:2003} and \citet{PreibischT:2005} who find a changed (from that for ZAMS stars) X-ray activity - Rossby number relationship for the pre-main sequence stars in the Orion nebula.

Like saturation, the factors causing supersaturation are still open to debate. \citet{StepienK:2001} have proposed that supersaturation is a result of the poleward migration of active coronal regions, leaving the equator void of active regions and thus decreasing the coronal filling factor of the star. The fact that the chromospheric emission observed in this paper shows little evidence of supersaturation would count against this model being the mechanism for supersaturation, as it would be expected that a poleward movement of active regions would also affect the chromospheric emission. 

Such a trend in the motion of active regions is also not evidenced in the Doppler imaging of spot topologies of young active stars. Many of the stars Doppler imaged show rather similar spot patterns often with polar or high latitude spots and some lower latitude features. Comparing the Doppler images of stars from IC 2391/2602 in both the saturation (R58, \vsinis = 92 \kmsn) and supersaturation (VXR45A, \vsinis = 235 \kmsn) regime show similar spot maps \citep{MarsdenSC:2004} with no evidence of a further poleward migration of active regions for the supersaturated star.

The coronal stripping model of \citet{JardineM:1999} discussed previously could account for supersaturation. As the co-rotation radius continues to decrease with an increase in rotation rate, eventually enough of the coronal volume is forced open and the X-ray emission falls with rotation rate. \citet{JardineM:2004} has used this model to explain both the modulation (observed by \citealt{MarinoA:2003}) and supersaturation of the X-ray observations of the ultra-rapid rotator VXR45A. If the coronal stripping model is correct then it would be unlikely that chromospheric emission would show any evidence of supersaturation as centrifugal stripping will only affect the coronae of these stars. Thus the observations presented in this paper, showing no chromospheric supersaturation, are consistent with the coronal stripping model of \citet{JardineM:1999}. Unfortunately, as mentioned previously, the fact that both the chromospheric and coronal emission show the onset of saturation at a similar Rossby number would appear to count against this model.

Thus the mechanisms causing both saturation and supersaturation remain elusive.

\section{Conclusions}\label{irtconclusions}

In this paper we have reported on the chromospheric emission levels of over 50 solar-type F, G, and K stars in the young open clusters IC 2391 and IC 2602. Due to their youth, IC 2391 and IC 2602 contain a number of star that can be classified as ultra-rapid rotators (\vsinis $>$ 100 \kmsn) and thus are excellent targets to determine if chromospheric emission shows signs of supersaturation. 

We have redetermined \vsinis values for all our targets in both clusters and determined \vsinis values for a number of targets (especially those ultra-rapid rotators) for the first time.  The clusters show a wide range of \vsinis values  for most colours studied (0.4 $<$ (V-I\subs{C})\subs{0} $<$ 1.4)  although there is a deficit of rapid rotators around (V-I\subs{C})\subs{0} $\sim$ 0.8, which we believe has just occurred by chance.

We have shown that chromospheric emission saturates for stars with Log(N\subs{R}) $\la$ -1.1. The saturation levels found are Log(R$^{\prime}$\subs{8542}) $\sim$ -4.2, Log(R$^{\prime}$\subs{8498}) $\sim$ -4.4, and Log(R$^{\prime}$\subs{8662}) $\sim$ -4.3, for the three Calcium IRT lines at 8542\AA, 8498\AA, and 8662\AA\/ respectively. The chromospheric saturation level in the 8542\AA\/ line agrees well with that for two older clusters, NGC 6475 and the Pleiades \citep{JamesDJ:1997,SoderblomDR:1993}.  Both activity indicators appear to be saturating at a similar Rossby number which lends weight to the saturation seen being caused by a single mechanism, which remains unknown.

Significantly, chromospheric emission from these stars shows little evidence of the effect of supersaturation seen in the coronal X-ray emission of the same stars. Thus we believe that X-ray supersaturation is not a result of an overall decrease in the efficiency of the magnetic dynamo in ultra-rapid rotators.

\section*{Acknowledgments}

The observations for this paper were taken with Anglo-Australian Telescope. We would like to thank the technical staff of the Anglo-Australian Observatory for their excellent assistance during these observations. We would also like to thank the referees (Rob Jeffries and an anonymous referee) for their comments that helped greatly improve this manuscript. SCM was funded by a University of Southern Queensland PhD scholarship during the course of this work. This research has made use of the WEBDA database, operated at the Institute for Astronomy of the University of Vienna.



\label{lastpage}


\begin{thebibliography}{}
\bibitem[\protect\citeauthoryear{Barnes}{2003}]{BarnesS:2003} Barnes S., 2003, ApJ, 586, 464
\bibitem[\protect\citeauthoryear{Barnes \& Sofia}{1996}]{BarnesS:1996} Barnes S., Sofia S., 1996, ApJ, 462, 746
\bibitem[\protect\citeauthoryear{Barnes et al.}{1999}]{BarnesSA:1999} Barnes S. A., Sofia S., Prosser C. F., Stauffer J. R., 1999, ApJ, 516, 263
\bibitem[\protect\citeauthoryear{Barrado y Navascu\'es, Stauffer, \& Jayawardhana}{Barrado y Navascu\'es et al.}{2004}]{BarradoD:2004} Barrado y Navascu\'es D., Stauffer J. R., Jayawardhana R., 2004, ApJ, 614, 386
\bibitem[\protect\citeauthoryear{Bessel \& Stringfellow}{1993}]{BessellMS:1993} Bessell M. S., Stringfellow G. S., 1993, in Burbidge G., Layzer D., Sandage A., eds., Annual Review of Astronomy and Astrophysics, Vol. 31, Annual Reviews Inc., Palo Alto, California, p. 433
\bibitem[\protect\citeauthoryear{Bessell, Castelli \& Plez}{Bessell et al.}{1998}]{BessellMS:1998} Bessell M. S., Castelli F., Plez B., 1998, A\&A, 333, 231
\bibitem[\protect\citeauthoryear{Bouvier, Forestini \& Allain}{Bouvier et al.}{1997}]{BouvierJ:1997} Bouvier J., Forestini M., Allain S., 1997, A\&, 326, 1023
\bibitem[\protect\citeauthoryear{Braes}{1961}]{BraesLLE:1961} Braes L. L. E., 1961, Monthly Notices of the Astronomical Society of South Africa, 20, 7
\bibitem[\protect\citeauthoryear{Braes}{1962}]{BraesLLE:1962} Braes L. L. E., 1962, Bulletin of the Astronomical Institutes of the Netherlands, 16, 297
\bibitem[\protect\citeauthoryear{Buscombe}{1965}]{BuscombeW:1965} Buscombe W., 1965, MNRAS, 129, 411
\bibitem[\protect\citeauthoryear{Caldwell et al.}{1993}]{CaldwellJAR:1993} Caldwell J.A.R., Cousins A.W.J., Ahlers C.C., van Wamelan P., Maritz E.J., 1993, South African Astronomical Observatory Circulars, 15, 1
\bibitem[\protect\citeauthoryear{Cardini \& Cassatella}{2007}]{CardiniD:2007} Cardini D., Cassatella A., 2007, ApJ, 666, 393
\bibitem[\protect\citeauthoryear{Chmielewski}{2000}]{ChmielewskiY:2000} Chmielewski Y., 2000, A\&A, 353, 666
\bibitem[\protect\citeauthoryear{D'Antona \& Mazzitelli}{1997}]{DAntonaF:1997} D'Antona F., Mazzitelli I., 1997, in Micela G., Pallavicini R., Sciortino S., eds., Memorie della Societa astronomia Italiana, Vol. 68, Cool stars in clusters and associations: Magnetic activity and age indicators, Societa astronomia Italiana, p.807
\bibitem[\protect\citeauthoryear{Donati et al.}{1997}]{DonatiJF:1997b} Donati J.-F., Semel M., Carter B., Rees D. E., Cameron A. C., 1997, MNRAS, 291, 658
\bibitem[\protect\citeauthoryear{Donati et al.}{2003}]{DonatiJF:2003} Donati J.-F., Collier Cameron A., Semel M., et al., 2003, MNRAS, 345, 1145
\bibitem[\protect\citeauthoryear{D'Orazi \& Randich}{2009}]{DOraziV:2009} D'Orazi V., Randich S., 2009, A\&A, accepted (arXiv:0905.1835)
\bibitem[\protect\citeauthoryear{Doyle}{1996}]{DoyleJG:1996} Doyle J. G., 1996, A\&A, 307, L45
\bibitem[\protect\citeauthoryear{Feigelson et al.}{2003}]{FeigelsonED:2003} Feigelson E. D., Gaffney III J. A., Garmire G., Hillenbrand L. A., Townsley L., 2003, ApJ, 584, 911
\bibitem[\protect\citeauthoryear{Foing et al.}{1989}]{FoingBH:1989} Foing B. H., Crivellari L., Vladilo G., Rebolo R., Beckman J. E., 1989, A\&A Supp. Ser., 80, 189
\bibitem[\protect\citeauthoryear{Hogg}{1960}]{HoggAR:1960} Hogg A. R., 1960, PASP, 72, 85
\bibitem[\protect\citeauthoryear{Irwin et al.}{2007}]{IrwinJ:2007} Irwin J., Hodgkin S., Aigrain S., Hobb L., Bouvier J., Clarke C., Moraux E., Bramich D. M., 2007, MNRAS, 377, 741
\bibitem[\protect\citeauthoryear{James \& Jeffries}{1997}]{JamesDJ:1997} James D. J., Jeffries R.D., 1997, MNRAS, 291, 252
\bibitem[\protect\citeauthoryear{James et al.}{2000}]{JamesDJ:2000} James D. J., Jardine M. M., Jeffries R. D., Randich S., Collier Cameron A., Ferreira M., 2000, MNRAS, 318, 1217
\bibitem[\protect\citeauthoryear{Jardine}{2004}]{JardineM:2004} Jardine M., 2004, A\&A, 414, L5
\bibitem[\protect\citeauthoryear{Jardine \& Unruh}{1999}]{JardineM:1999} Jardine M., Unruh Y. C., 1999, A\&A, 346, 883
\bibitem[\protect\citeauthoryear{Jeffries \& Oliveira}{2005}]{JeffriesRD:2005} Jeffries R. D., Oliveira J. M., 2005, MNRAS, 358, 13
\bibitem[\protect\citeauthoryear{Jeffries, Totten, \& James}{Jeffries et al.}{2000}]{JeffriesRD:2000} Jeffries R. D., Totten E. J., James D. J., 2000, MNRAS, 316, 950
\bibitem[\protect\citeauthoryear{Kim \& Demarque}{1996}]{KimYC:1996} Kim Y.-C., Demarque P., 1996, ApJ, 457, 340
\bibitem[\protect\citeauthoryear{Krishnamurthi et al.}{1997}]{KrishnamurthiA:1997} Krishnamurthi A., Pinsonneault M. H., Barnes S., Sofia S., 1997, ApJ, 480, 303
\bibitem[\protect\citeauthoryear{Krishnamurthi et al.}{1998}]{KrishnamurthiA:1998} Krishnamurthi A., Terndrup D. M., Pinsonneault M. H., et al., 1998, ApJ, 493, 914
\bibitem[\protect\citeauthoryear{Kurucz}{1993}]{KuruczRL:1993} Kurucz R. L., 1993, CDROM \#13 (ATLAS9 atmospheric models) and CDROM \#18 (ATLAS9 and SYNTHE routines, spectral line database)
\bibitem[\protect\citeauthoryear{Linsky et al.}{1979}]{LinskyJL:1979} Linsky J. L., Hunten D. M., Sowell R., Glackin D. L., Kelch W. L., 1979, ApJ Supp. Ser., 41, 481
\bibitem[\protect\citeauthoryear{Lyng\.a}{1961}]{LyngaG:1961} Lyng\.a G., 1961, Arkiv f\"or Astronomii, 2, 379
\bibitem[\protect\citeauthoryear{Mallik}{1997}]{MallikSV:1997} Mallik S. V., 1997, A\&A Supp. Ser., 124, 359
\bibitem[\protect\citeauthoryear{Marino et al.}{2003}]{MarinoA:2003} Marino A.,  Micela G., Peres G., Sciortino S., 2003, A\&A, 407, L63
\bibitem[\protect\citeauthoryear{Marino et al.}{2005}]{MarinoA:2005} Marino A.,  Micela G., Peres G., Pillitteri I., Sciortino S., 2005, A\&A, 430, 287
\bibitem[\protect\citeauthoryear{Marsden et al.}{2004}]{MarsdenSC:2004} Marsden S. C., Waite I. A., Carter B. D., Donati J.-F., 2004, Astron. Nachr., 325, 246
\bibitem[\protect\citeauthoryear{Marsden et al.}{2006}]{MarsdenSC:2006} Marsden S. C., Donati J.-F., Semel M., Petit P., Carter B. D., 2006, MNRAS, 370, 468
\bibitem[\protect\citeauthoryear{Mestel \& Spruit}{1987}]{MestelL:1987} Mestel L., Spruit H. C., 1987, MNRAS, 226, 57
\bibitem[\protect\citeauthoryear{Micela et al.}{1999}]{MicelaG:1999} Micela G., Sciortino S., Harnden Jr F.R., et al., 1999, A\&A, 341, 751
\bibitem[\protect\citeauthoryear{Nordstr\"{o}m et al.}{2004}]{NordstromB:2004} Nordstr\"{o}m B., Mayor M., Anderson J., et al., 2004, A\&A, 418, 989
\bibitem[\protect\citeauthoryear{Noyes et al.}{1984}]{NoyesRW:1984} Noyes R. W., Hartmann L. W., Baliunas S. L., Duncan D. K., Vaughan A. H., 1984, ApJ, 279, 763
\bibitem[\protect\citeauthoryear{O'Dell et al.}{1995}]{ODellMA:1995} O'Dell M. A., Panagi P., Hendry M. A., Collier Cameron A., 1995, A\&A, 294, 715
\bibitem[\protect\citeauthoryear{Patten}{1995}]{PattenBM:1995} Patten B. M., 1995, PhD thesis, University of Hawaii
\bibitem[\protect\citeauthoryear{Patten \& Simon}{1996}]{PattenBM:1996} Patten B. M., Simon T., 1996, ApJ, Supp. Ser., 106, 489
\bibitem[\protect\citeauthoryear{Perry \& Bond}{1969}]{PerryCL:1969b} Perry C. L., Bond H. E., 1969, PASP, 81, 629
\bibitem[\protect\citeauthoryear{Perry \& Hill}{1969}]{PerryCL:1969a} Perry C. L., Hill G., 1969, AJ, 74, 899
\bibitem[\protect\citeauthoryear{Perryman et al.}{1998}]{PerrymanMAC:1998} Perryman M.A.C., Brown A.G.A., Lebreton Y., et al., 1998, A\&A, 331, 81
\bibitem[\protect\citeauthoryear{Pinsonneault et al.}{1998}]{PinsonneaultMH:1998} Pinsonneault M. H., Stauffer J., Soderblom D. R., King J. R., Hanson J. B., 1998, ApJ, 504, 170
\bibitem[\protect\citeauthoryear{Pizzolato et al.}{2003}]{PizzolatoN:2003} Pizzolato N., Maggio A., Micela G., Sciortino S., Ventura P., 2003, A\&A, 397, 147
\bibitem[\protect\citeauthoryear{Platais et al.}{2007}]{PlataisI:2007} Platais I., Melo C., Mermilliod J.-C., Kozhurina-Platais V., Fulbright J. P., M\'endez R. A., Altmann M., Sperauskas J., 2007, A\&A, 461, 509
\bibitem[\protect\citeauthoryear{Preibisch et al.}{2005}]{PreibischT:2005} Preibisch T., Kim Y.-C., Fatava F., et al., 2005, ApJ Supp. Ser., 160, 401
\bibitem[\protect\citeauthoryear{Prosser, Randich \& Stauffer}{Prosser et al.}{1996a}]{ProsserCF:1996a} Prosser C. F., Randich S., Stauffer J. R., 1996a, AJ, 112, 649
\bibitem[\protect\citeauthoryear{Prosser et al.}{1996b}]{ProsserCF:1996b} Prosser C. F., Randich S., Stauffer J. R., Schmitt J. H. M. M., Simon T., 1996b, AJ, 112, 1570
\bibitem[\protect\citeauthoryear{Radick et al.}{1987}]{RadickRR:1987} Radick R.R., Thompson D.T., Lockwood G.W., Duncan D.K., Baggett W.E., 1987, ApJ, 321, 459
\bibitem[\protect\citeauthoryear{Randich}{1998}]{RandichS:1998} Randich S., 1998, in Donahue R.A., Bookbinder J.A., eds., Cool Stars, Stellar Systems and the Sun 10, ASP Conference Series, Vol.154, p.501
\bibitem[\protect\citeauthoryear{Randich}{2001}]{RandichS:2001b} Randich S., 2001, A\&A, 377, 512
\bibitem[\protect\citeauthoryear{Randich et al.}{1995}]{RandichS:1995} Randich S., Schmitt J. H. M. M., Prosser C. F., Stauffer J. R., 1995, A\&A, 300, 134
\bibitem[\protect\citeauthoryear{Randich et al.}{1996}]{RandichS:1996} Randich S., Schmitt J. H. M. M., Prosser C. F., Stauffer J. R., 1996, A\&A, 305, 785
\bibitem[\protect\citeauthoryear{Randich et al.}{1997}]{RandichS:1997a} Randich S., Aharpour N., Pallavicini R., Prosser C. F., Stauffer J. R., 1997, A\&A, 323, 86
\bibitem[\protect\citeauthoryear{Randich et al.}{2001}]{RandichS:2001a} Randich S., Pallavicini R., Meola G., Stauffer J. R., Balachandran S. C., 2001, A\&A, 372, 862
\bibitem[\protect\citeauthoryear{Scholz et al.}{2007}]{ScholzA:2007} Scholz A., Coffey J., Brandeker A., Jayawardhana R., 2007, ApJ, 662, 1254
\bibitem[\protect\citeauthoryear{Soderblom et al.}{1993}]{SoderblomDR:1993} Soderblom D. R., Stauffer J. R., Hudon J. D., Jones B. F., 1993, ApJ Supp. Ser., 85, 315
\bibitem[\protect\citeauthoryear{Stauffer et al.}{1989}]{StaufferJR:1989} Stauffer J., Hartmann L. W., Jones B. F., McNamara B. R., 1989, ApJ, 342, 285
\bibitem[\protect\citeauthoryear{Stauffer et al.}{1994}]{StaufferJR:1994} Stauffer J.R., Caillault J.-P., Gang\'{e} M., Prosser C.F., Hartmann L.W., 1994, ApJ Supp. Ser., 91, 625
\bibitem[\protect\citeauthoryear{Stauffer et al.}{1997}]{StaufferJR:1997} Stauffer J. R., Hartmann L. W., Prosser C. F., Randich S., Balachandran S., Patten B. M., Simon T., Giampapa M., 1997, ApJ, 479, 776
\bibitem[\protect\citeauthoryear{St\c epie\'n, Schmitt, \& Voges}{St\c epie\'{n} et al.}{2001}]{StepienK:2001} St\c epie\'n K., Schmitt J. H. M. M., Voges W., 2001, A\&A, 370, 157
\bibitem[\protect\citeauthoryear{Valenti \& Fischer}{2005}]{ValentiJA:2005} Valenti J. A., Fischer D. A., 2005, ApJ Supp. Ser., 159, 141
\bibitem[\protect\citeauthoryear{van Leeuwen}{1999}]{vanLeeuwenF:1999} van Leeuwen F., 1999, A\&A, 341, L71
\bibitem[\protect\citeauthoryear{Whiteoak}{1961}]{WhiteoakJB:1961} Whiteoak J. B., 1961, MNRAS, 123, 245
\end{thebibliography}
\end{document}


\begin{landscape}
\begin{table*}
\begin{center}
\begin{minipage}{240mm}
\caption{Physical parameters of the solar-type stars observed in IC 2391. The first four columns give the star name, date of observation, number of observations, and signal-to-noise (around the 8542\AA\/ line) of the combined observations. Columns 5, 6, \& 7 list the photometry with \sups{a}: (B-V)\subs{0} estimated from (V-I\subs{C})\subs{0}, and \sups{b}: (V-I\subs{C})\subs{0} estimated from (B-V)\subs{0} from the formulas of Caldwell et al.\ (1993). Column 8 gives the effective temperature and Column 9 is the spectral type (from Patten \& Simon 1996). Column 10 is the strength of the Lithium line (N: No line evident, W: Weaker than the Ca 6717\AA\/ line, S: Similar or Stronger than Ca 6717\AA). Columns 11, 12, 13, \& 14 give the rotational period (\sups{c}: represents an uncertain value), the radial velocity from this study, the \vsinis from this study, and the \vsinis from previous works. Column 15 is a note on the LSD profiles, A: some Asymmetry in the LSD profile, M: Multiple star, E: larger Error of $\pm$5 \kmss in both Radial and Rotational Velocities, and P: Possible Binary. Column 16 has our cluster membership determination (M: Member, N: Non-member, and ?: probable member), and Columns 17, 18, \& 19 give the stars Luminosity, mass, and radius (only calculated for single members of the cluster).} \label{irt2391rot} 
\scriptsize
\begin{tabular}{lcccccccccccccccccc}
\hline
Star   & UT Obs.     & \# & S/N & V     & (B-V)\subs{0} & (V-I\subs{C})\subs{0} & T\subs{eff} & Sp. & Li & Period & Rad. Vel. & \vsini\subs{ours} & \vsini\subs{previous} & Notes & Member & Log(L/\Lsolar) & Mass      & Radius    \\
Name   & Date        & Obs. &       & (mag) & (mag)         & (mag)                 & (K)         & Type &     & (d) & (\kmsn)         & (\kmsn)                  & (\kmsn)                      &       & ?      &                & (\Msolar) & (\Rsolar) \\
\hline
H21    & 08 Jan 2001 & 2 & 40 & 11.69 & 0.81          & 0.86\sups{b}          & 5310        &  - & N  &        & 53              & $\le$7	              &                              &       & N      &        &       &        \\
H30    & 08 Jan 2001 & 2 & 45 & 10.89 & 0.83          & 0.88\sups{b}          & 5250        & - & N  &        & 19              & 18                       &                              &       & N      &          &        &       \\
L32    & 08 Jan 2001 &  2 & 48 & 9.38 & 0.43          & 0.51\sups{b}          & 6720        & F3V & S  &        & 16              & 68	                      &                              &       & M      & 0.6472         & 1.5       & 1.6       \\
SHJM7  & 12 Feb 2001 & 2 & 48 & 12.52 & 1.09          & 1.20                  & 4520        & G7 & W  &        & 6               & $\le$7                   &                              &       & N      &       &       &       \\
VXR02A & 08 Jan 2001 & 2 & 34 & 11.55 & 0.90\sups{a}  & 0.97                  & 5020        & G9 & S  &        & {\bf 16}              & 235                      &                              & P,E   & M      & -0.2686        & 1.0       & 1.0       \\
VXR02B & 12 Feb 2001 & 2 & 44 & 12.74 & 1.11\sups{a}  & 1.23                  & 4470        & K3 & W  &        & 16              & 8                        &                              &       & N      &        &       &       \\
VXR03A & 08 Jan 2001 & 2 & 45 & 10.95 & 0.67\sups{a}  & 0.73                  & 5740        & - & S  &        & 16              & 10                       & 10                           &       & M      & -0.0985        & 1.0       & 0.9       \\
VXR05  & 07 Jan 2001 & 2 & 45 & 10.29 & 0.67          & 0.50                  & 6670        & G2 & S  &        & 41/10           & $\le$7/$\le$7            & 34                           & M     & ?      &          &       &        \\
VXR07  & 08 Jan 2001 & 2 & 45 & 9.63 & 0.45          & 0.53\sups{b}          & 6630        & F5V & S  &        & 12              & 21                       & 21                           &       & ?      & 0.5553         & 1.4       & 1.4       \\
VXR08  & 06 Jan 2001 & 2 & 76 & 8.78 & 0.40          & 0.48\sups{b}          & 6870        & F3V & W  &        & 12              & 170                      &                              & E     & N      &          &       &        \\
VXR11    & 06 Jan 2001 &  2 & 57 & 9.96 & 0.63          & 0.69\sups{b}          & 5900        & - & S  &        & -94/45/90       & 14/8/14                  &                              & M     & ?      &         &       &       \\
VXR12  & 11 Feb 2001 & 4 & 48 & 11.86 & 0.83          & 0.90                  & 5210        & K0 & S  & 3.86   & 16              & 10                       & 16                           &       & M      & -0.4153        & 0.9       & 0.8       \\
VXR14  & 06 Jan 2001 & 2 & 64 & 10.45 & 0.56          & 0.68                  & 5920        & G0 & S  & 1.32   & 17              & 43                       & 47                           &       & M      & 0.0901         & 1.1       & 1.1       \\
VXR16A & 11 Feb 2001 & 3 & 38 & 11.84 & 0.87\sups{a}  & 0.93                  & 5130        & G9 & S  &        & {\bf 15}              & 21                       & 22                           & A     & M      & -0.3978        & 0.9       & 0.8       \\
VXR22A & 08 Jan 2001 & 2 & 45 & 11.08 & 0.73\sups{a}  & 0.79                  & 5540        & G1 & S  &        & 15              & 8                        &                              &       & M      & -0.1353        & 1.0       & 0.9       \\
VXR30  & 07 Jan 2001 &  2 & 54 & 9.83 & 0.48          & 0.56\sups{b}          & 6500        & F6V & S  &        & 36              & 41                       & 43                           &       & N      &       &       &       \\
VXR31  & 08 Jan 2001 & 2 & 33 & 11.22 & 0.66\sups{a}  & 0.72                  & 5770        & - & S  &        & 24              & 17                       & 17                           &       & ?      & -0.2089        & 1.0       & 0.8       \\
VXR35A & 07 Jan 2001 & 3 & 61 & 12.63 & 0.99          & 1.24                  & 4450        & K3e & S  & 0.527  & 16              & 89                       & 90                           &       & M      & -0.5987        & 0.8       & 0.8       \\
VXR44  & 06 Jan 2001 &  2 & 68 & 9.69 & 0.41          & 0.49\sups{b}          & 6820        & F6V & S  &        & 12              & 79                       & 67                           &       & ?      & 0.5156         & 1.4       & 1.3       \\
VXR45A & 06 Jan 2001 & 2 & 69 & 10.70 & 0.80          & 0.89                  & 5240        & G9 & S  & 0.223  & 16              & 235                      & $>$ 200                      & E     & M      & 0.0456         & 1.2       & 1.3       \\
VXR50A & 11 Feb 2001 & 3 & 50 & 12.54 & 0.84\sups{a}  & 0.90                  & 5210        & G6 & S  &        & {\bf 22}              & 56                       & 64                           &       & ?      & -0.6873        & 0.8       & 0.6       \\
VXR52  & 07 Jan 2001 & 2 & 49 & 10.34 & 0.56          & 0.64                  & 6080        & F9 & S  &        & 16              & 10                       & $<$ 15                       &       & M      & 0.1259         & 1.2      & 1.0       \\
VXR62A & 08 Jan 2001 & 4 & 58 & 11.73 & 0.85          & 0.98                  & 5000        & K0 & S  & 0.503\sups{c}  & 14              & 49                       & 52                           &       & ?      & -0.3372        & 0.9       & 0.9       \\
VXR66  & 07 Jan 2001 &  2 & 55 & 9.73 & 0.46          & 0.55                  & 6450        & F5 & S  &        & 15              & 52                       &                              &       & M      & 0.3548         & 1.3       & 1.2       \\
VXR67A & 11 Feb 2001 & 3 & 29 & 11.71 & 0.95\sups{a}  & 1.02                  & 4900        & K0 & S  &        & 19              & 9                        & 8                            & A     & ?      & -0.3153        & 1.0       & 1.0       \\
VXR69A & 11 Feb 2001 & 4 & 32 & 11.67 & 0.83\sups{a}  & 0.89                  & 5240        & K0 & S  &        & 16              & 19                       & 19                           &       & M      & -0.3424        & 0.9       & 0.8       \\
VXR70  & 08 Jan 2001 & 2 & 33 & 10.85 & 0.63          & 0.74                  & 5700        & G3 & S  & 2.67   & 13              & 16                       & 17                           &       & ?      & -0.0561        & 1.0       & 1.0       \\
VXR72  & 11 Feb 2001 & 3 & 45 & 11.46 & 0.72          & 0.83                  & 5410        & G9 & S  & 3.23   & 15              & 13                       & 15                           &       & M      & -0.2763        & 0.9       & 0.8       \\
VXR76A & 12 Feb 2001 & 2 & 49 & 12.76 & 1.04          & 1.23                  & 4470        & K3 & S  & 4.86   & 15              & $\le$7                   & 8                            &       & M      & -0.6548        & 0.8       & 0.8       \\
VXR77A & 06 Jan 2001 &  3 & 57 & 9.91 & 0.49          & 0.59                  & 6270        & F8 & S  & 0.653  & 17              & 93                       & 95                           &       & M      & 0.2889         & 1.3       & 1.2       \\
VXR78  & 06 Jan 2001 & 2 & 79 & 10.44 & 0.66\sups{a}  & 0.72                  & 5770        & G3 & S  &        & 23              & 46                       & 50                           &       & ?      & 0.1031         & 1.1       & 1.1       \\
VXR80A & 06 Jan 2001 & 4 & 60 & 11.98 & 0.96\sups{a}  & 1.03                  & 4880        & K2 & S  &        & 17              & 145                      & 150                          & E     & M      & -0.4197        & 0.9       & 0.9       \\
\hline
\end{tabular}
\end{minipage}
\end{center}
\end{table*}
\end{landscape}																																										
\begin{landscape}
\begin{table*}
\begin{center}
\begin{minipage}{240mm}
\caption{Physical parameters of the solar-type stars observed in IC 2602. The columns are the same as in Table~\ref{irt2391rot}, except that the spectral type comes from \sups{d}: Patten (1995), \sups{e}: Braes (1962), and \sups{f}: Whiteoak (1961).} 
\label{irt2602rot}
\scriptsize
\begin{tabular}{lcccccccccccccccccc}
\hline
Star & UT Obs.     & \# & S/N & V     & (B-V)\subs{0} & (V-I\subs{C})\subs{0} & T\subs{eff} & Sp. & Li & Period       & Rad. Vel. & \vsini\subs{ours} & \vsini\subs{previous} & Notes & Member & Log(L/\Lsolar) & Mass      & Radius    \\
Name & Date        & Obs. &        & (mag) & (mag)         & (mag)                 & (K)         & Type &    & (d)       & (\kmsn)         & (\kmsn)                  & (\kmsn)                      &       & ?      &                & (\Msolar) & (\Rsolar) \\    
\hline
B56  & 12 Feb 2001 & 2 & 52 & 10.26 & 0.54          & 0.71                  & 5810        & G0\sups{e} & N  &              & 19              & {\bf $\le$11}                   & $<$ 7                        &       & N      &          &       &       \\
B57  & 12 Feb 2001 & 2 & 52 & 9.98 & 0.54          & 0.62                  & 6150        & F8V\sups{f} & W  &              & 19              & {\bf $\le$11}                   & $<$ 7                        &       & N      &        &       &       \\
B71  & 19 Mar 2000 & 2 & 42 & 10.88 & 0.56          & 0.65                  & 6030        & G0\sups{f} & S  &              & 5/9             & {\bf $\le$11/$\le$11}     &                              & M     & ?      &       &       &       \\
B79  & 19 Mar 2000 & 2 & 58 & 10.10 & 0.50          & 0.58                  & 6320        & A5\sups{e} & N  &              & {\bf -18/56}          & {\bf $\le$11}/45          &                              & M,E   & N      &         &        &        \\
B90  & 11 Feb 2001 & 3 & 42 & 10.82 & 0.69          & 0.92                  & 5150        & A3\sups{e} & N  &              & -78             & 57                       &                              & P     & N      &         &        &        \\
B104 & 17 Mar 2000 & 2 & 65 & 9.98 & 0.60          & 0.67                  & 5960        & F5\sups{f} & N  &              & {\bf -1}               & 175                      & $>$ 150                      & E     & N      &          &        &        \\
B112 & 12 Feb 2001 & 2 & 61 & 10.93 & 0.81          & 0.95                  & 5070        & G5\sups{e} & N  &              & 38              & $\le$7                   &                              &       & N      &         &       &       \\
B113 & 19 Mar 2000 & 2 & 45 & 10.94 & 0.72          & 0.80                  & 5500        & F5\sups{e} & W  &              & ?/?             & ?/?                      &                              & M     & N      &        &       &       \\
B116 & 12 Feb 2001 & 2 & 52 & 10.74 & 0.86          & 0.96                  & 5050        & G0\sups{e} & N  &              & 19              & 39                       &                              &       & N      &         &       &       \\
B119 & 18 Mar 2000 & 2 & 66 & 10.84 & 0.59          & 0.75                  & 5670        & F2\sups{e} & N  &              & 6               & 10                       &                              & A,E   & N      &         &       &       \\
B128 & 12 Feb 2001 & 2 & 71 & 9.62 & 0.53          & 0.56                  & 6400        & A5\sups{e} & N  &              & {\bf -1}               & 215                      &                              & E     & N      &          &        &       \\
B140 & 19 Mar 2000 & 2 & 52 & 10.77 & 0.71          & 0.84                  & 5380        & F2\sups{e} & N  &              & {\bf 65}              & 30                       &                              &       & N      &        &       &       \\
B143 & 12 Feb 2001 & 2 & 54 & 10.08 & 0.44          & 0.50                  & 6670        & A3\sups{e} & N  &              & {\bf -11}             & 170                      &                              & E     & N      &         &       &       \\
R1   & 18 Mar 2000 & 3 & 64 & 11.57 & 0.87          & 0.86                  & 5320        & - & S  &              & {\bf 18}              & $\le$7                   &                              &       & M      & -0.2995        & 0.9       & 0.8       \\
R2   & 12 Feb 2001 & 2 & 55 & 12.52 & 1.21          & 1.41                  & 4180        & - & W  &              & 0               & 120                      &                              & P,E   & N      &         &        &        \\
R3A  & 17 Mar 2000 & 2 & 65 & 11.32 & 0.83          & 0.84                  & 5380        & - & S  &              & 17/19           & 24/21                    & 25                           & M     & ?      &        &       &       \\
R4   & 19 Mar 2000 & 2 & 68 & 12.19 & 1.23          & 1.33                  & 4300        & - & N  &              & {\bf -7/17/76}        & $\le$7/$\le$7/$\le$7     &                              & M     & N      &        &        &        \\
R7   & 17 Mar 2000 & 2 & 99 & 9.21 & 0.40          & 0.41                  & 7120        & - & S  &              & 17              & 58                       & 52                           &       & M      & 0.5613         & 1.5       & 1.3       \\
R8A  & 18 Mar 2000 & 2 & 72 & 10.41 & 0.61          & 0.61                  & 6190        & - & S  &              & 17              & 27                       &                              &       & M      & 0.1043         & 1.2       & 1.0       \\
R10  & 12 Feb 2001 & 2 & 28 & 12.77 & 1.19          & 1.20                  & 4520        & - & S  &              & 19              & 14                       &                              &       & M      & -0.6590        & 0.8       & 0.8       \\
R14  & 18 Mar 2000 & 3 & 57 & 11.57 & 0.83          & 0.83                  & 5410        & - & S  &              & {\bf 16}              & 11                       & 13                           &       & M      & -0.3083        & 0.9       & 0.8       \\
R15A & 18 Mar 2000 & 3 & 73 & 11.75 & 0.89          & 1.01                  & 4920        & - & S  & 3.6          & 17              & $\le$7                   & 10                           &       & M      & -0.3228        & 1.0       & 0.9       \\
R18  & 18 Mar 2000 & 2 & 61 & 12.49 & 1.01          & 1.15                  & 4620        & - & N  &              & 64/11           & 95/$\le$7                & 11                           & M,E   & N      &        &        &        \\
R21  & 17 Mar 2000 & 2 & 82 & 9.50 & 0.47          & 0.57                  & 6360        & G0\sups{e} & S  &              & 19/15           & 39/?                     & 23                           & M     & ?      &         &        &        \\
R29  & 19 Mar 2000 & 3 & 69 & 12.73 & 1.07          & 1.14                  & 4640        & K4\sups{d} & S  & 2.21         & 18              & 21                       & 22                           &       & M      & -0.6667        & 0.8       & 0.7       \\
R35  & 17 Mar 2000 & 2 & 69 & 10.59 & 0.36          & 0.66                  & 5990        & G5\sups{f} & S  &              & 17              & 21                       & 20                           &       & M      & 0.0419         & 1.1       & 1.0       \\
R42C & 19 Mar 2000 & 3 & 64 & 11.56 & 0.76          & 0.82                  & 5440        & - & S  &              & {\bf 19}              & 10                       &                              &       & M      & -0.3071        & 0.9       & 0.8       \\
R43  & 17 Mar 2000 & 3 & 62 & 12.14 & 0.91          & 1.05                  & 4830        & K0\sups{d} & S  & 0.78         & 16              & 47                       & 50                           &       & M      & -0.4645        & 0.9       & 0.8       \\
R45A & 17 Mar 2000 & 2 & 75 & 10.73 & 0.62          & 0.67                  & 5960        & - & S  &              & 18              & 14                       & 23                           &       & M      & -0.0120        & 1.1       & 0.9       \\
R46  & 18 Mar 2000 & 2 & 79 & 10.70 & 0.63          & 0.77                  & 5600        & - & W  &              & -13/63          & ?/17                     &                              & M     & N      &         &        &        \\
R52  & 17 Mar 2000 & 4 & 62 & 12.19 & 1.03          & 1.06                  & 4810        & K4\sups{d} & S  & 0.39         & 14              & 122                      & 95                           &       & ?      & -0.4809        & 0.9       & 0.8       \\
R54A & 12 Feb 2001 & 2 & 58 & 12.13 & 1.11          & 1.32                  & 4320        & - & S  &              & 19/16           & 64/?                     &                              & M     & ?      &        &        &        \\
R58  & 17 Mar 2000 & 2 & 57 & 10.52 & 0.61          & 0.69                  & 5880        & G0\sups{e} & S  & 0.57         & 17              & 92                       & 93                           &       & M      & 0.0763         & 1.1       & 1.1       \\
R59  & 17 Mar 2000 & 3 & 64 & 11.86 & 0.78          & 0.95                  & 5070        & K0\sups{d} & S  & 1.31         & 19              & 31                       & 34                           &       & M      & -0.3873        & 0.9       & 0.8       \\
R66  & 18 Mar 2000 & 2 & 62 & 11.07 & 0.64          & 0.78                  & 5570        & - & S  & 3.3          & 17              & 11                       & 12                           &       & M      & -0.1220        & 1.0       & 0.9       \\
R68  & 17 Mar 2000 & 3 & 90 & 11.32 & 0.82          & 1.04                  & 4850        & K0\sups{d} & S  & 0.99\sups{c} & 19              & 51                       & 48                           &       & M      & -0.1401        & 1.1       & 1.2       \\
R70  & 18 Mar 2000 & 2 & 57 & 10.92 & 0.65          & 0.66                  & 5990        & F7\sups{f} & S  & 4.3          & 17              & 9                        & 10                           &       & M      & -0.0901        & 1.1       & 0.8       \\
R72  & 17 Mar 2000 & 2 & 65 & 10.89 & 0.60          & 0.71                  & 5810        & G0\sups{e} & S  & 1.05         & 18              & 49                       & 49                           &       & M      & -0.0672        & 1.1       & 0.9       \\
R79  & 17 Mar 2000 & 2 & 103 & 9.08 & 0.40          & 0.46                  & 6870        & F2\sups{f} & S  &              & 12              & 78                       & 57                           & A     & ?      & 0.6166         & 1.5       & 1.4       \\
R80  & 18 Mar 2000 & 2 & 66 & 10.66 & 0.89          & 0.98                  & 4980        & G5\sups{e} & S  & 6.7          & 19              & 13                       & 10                           &       & M      & 0.1028         & 1.3       & 1.5       \\
R83  & 17 Mar 2000 & 2 & 78 & 10.70 & 0.58          & 0.73                  & 5740        & - & S  & 1.67         & 17              & 30                       & 30                           &       & M      & 0.0135         & 1.0       & 1.0       \\
R85  & 17 Mar 2000 & 2 & 76 & 9.87 & 0.48          & 0.53                  & 6530        & F8\sups{e} & S  &              & {\bf 16}              & 45                       & 45                           &       & M      & 0.3080         & 1.3       & 1.1       \\
R88A & 17 Mar 2000 & 3 & 69 & 12.71 & 1.16          & 1.30                  & 4350        & K4\sups{d} & S  & 0.204        & 16              & 255                      & $>$ 200                      & E     & M      & -0.5937        & 0.8       & 0.9       \\
R89  & 19 Mar 2000 & 2 & 48 & 12.97 & 1.20          & 1.30                  & 4350        & K4\sups{d} & S  & 4.8          & 17              & 9                        & 14                           &       & M      & -0.6977        & 0.8       & 0.8       \\
R92  & 19 Mar 2000 & 2 & 102 & 10.26 & 0.63          & 0.73                  & 5740        & G\sups{e} & S  & 2.0\sups{c}  & 17              & 16                       & 14                           &       & M      & 0.1895         & 1.1       & 1.3       \\
R95A & 18 Mar 2000 & 3 & 52 & 11.73 & 0.83          & 0.92                  & 5150        & G9\sups{d} & S  & 1.20         & 16              & 14                       & 12                           &       & M      & -0.3450        & 0.9       & 0.8       \\
R96  & 12 Feb 2001 & 2 & 25 & 12.94 & 1.21          & 1.32                  & 4320        & K4\sups{d}& S  & 1.82         & 17              & 15                       & 17                           &       & M      & -0.6772        & 0.8       & 0.8       \\
W75  & 19 Mar 2000 & 3 & 54 & 10.44 & 0.66          & 0.72                  & 5770        & G0\sups{f} & W  &              & {\bf 1}               & $\le$7                   &                              &       & N      &          &       &       \\
W78  & 19 Mar 2000 & 2 & 47 & 10.94 & 0.55          & 0.69                  & 5880        & G0\sups{f} & W  &              & {\bf 3}               & {\bf $\le$11}                 &                              &       & N      &        &       &       \\
W79  & 18 Mar 2000 & 2 & 57 & 11.57 & 0.79          & 0.80                  & 5500        & G\sups{f} & S  & 6.2          & {\bf 18}              & 8                        & 8                            &       & M      & -0.3166        & 0.9       & 0.8       \\
W91  & 12 Feb 2001 & 2 & 49 & 10.73 & 0.74          & 0.93                  & 5130        & G0\sups{f} & S  &              & 3/?             & 13/?                     &                              & P     & ?      &        &       &       \\
\hline
\end{tabular}
\end{minipage}
\end{center}
\end{table*}
\end{landscape}

\setcounter{table}{3}

\begin{landscape}
\begin{table*}
\begin{center}
\begin{minipage}{240mm}
\caption{Chromospheric IRT emission of the single members observed in IC 2391. Column 1 is the star name and Column 2 is the Log of the Rossby Number for the star, where \sups{g}:
means that the Rossby Number was calculated from Equation 10. Column 3 is the Log of the X-ray emission for each star taken from Patten \& Simon (1996).
Columns 4, 8, \& 12 are the emission equivalent widths of the three IRT lines (8542\AA, 8498\AA, and 8662\AA). Columns 5, 9, \& 13 are the chromospheric emission flux in
each of the IRT lines. Columns 6, 10, \& 14 are the chromospheric emission ratios for each of the IRT lines and Columns 7, 11, \& 15 are the Logs of the chromospheric
emission ratios.} \label{irt2391chromo}
\scriptsize
\begin{tabular}{lcccccccccccccc}
\hline
Star & Log(N\subs{R}) & Log(L\subs{X}/L\subs{bol}) & W$^{\rm em}_{\rm 8542}$ & F$^{\prime}_{\rm 8542}$ (x10\sups{3}) & R$^{\prime}_{\rm 8542}$ & Log(R$^{\prime}_{\rm 8542}$) & W$^{\rm em}_{\rm 8498}$ & F$^{\prime}_{\rm 8498}$ (x10\sups{3}) & R$^{\prime}_{\rm 8498}$ & Log(R$^{\prime}_{\rm 8498}$) & W$^{\rm em}_{\rm 8662}$ & F$^{\prime}_{\rm 8662}$ (x10\sups{3}) & R$^{\prime}_{\rm 8662}$ & Log(R$^{\prime}_{\rm 8662}$) \\
Name &                &                            & (m\AA)                  & (Wm\sups{-2})                         & (x10\sups{-5})          &                              & (m\AA)                  & (Wm\sups{-2})                         & (x10\sups{-5})          &                              & (m\AA)                  & (Wm\sups{-2})                         & (x10\sups{-5})          &                              \\
\hline
L32         & -0.3401\sups{g} & $<$ -5.48 & 170 & 1.2 & 1.0 & -5.00 &  90 & 0.7 & 0.6 & -5.22 & 120 & 0.9 & 0.8 & -5.10 \\
VXR02A & -2.1128\sups{g} & -3.70        & 770 & 2.2 & 6.1 & -4.21 & 530 & 1.5 & 4.2 & -4.38 & 590 & 1.7 & 4.7 & -4.33 \\
VXR03A & -0.5636\sups{g} & -4.21        & 240 & 1.1 & 1.8 & -4.74 & 180 & 0.8 & 1.3 & -4.89 & 230 & 1.0 & 1.6 & -4.80 \\
VXR07   &  0.0107\sups{g} & -3.98         &  10 & 0.1 & 0.1 & -6.00 &  30    & 0.2 & 0.2 & -5.70 &  50 & 0.3 & 0.3 & -5.52 \\
VXR12   & -0.7242               & -3.63        & 510 & 1.7 & 4.1 & -4.39 & 360 & 1.2 & 2.9 & -4.54 & 430 & 1.4 & 3.4 & -4.47 \\
VXR14   & -0.7326               & -3.59        & 520 & 2.6 & 3.7 & -4.43 & 330 & 1.7 & 2.4 & -4.62 & 450 & 2.2 & 3.2 & -4.49 \\
VXR16A & -1.1475\sups{g} & -3.25       & 570 & 1.8 & 4.6 & -4.34 & 460 & 1.4 & 3.6 & -4.44 & 530 & 1.6 & 4.1 & -4.39 \\
VXR22A & -0.5945\sups{g} & -3.27      & 480 & 2.0 & 3.7 & -4.43 & 370 & 1.5 & 2.8 & -4.55 & 410 & 1.6 & 3.0 & -4.52 \\
VXR31   & -0.8249\sups{g} & -4.18      & 300 & 1.4 & 2.2 & -4.66 & 210 & 1.0 & 1.6 & -4.80 & 230 & 1.1 & 1.7 & -4.77 \\
VXR35A & -1.6392               & -3.00      & 820 & 1.4 & 6.5 & -4.19 & 460 & 0.8 & 3.7 & -4.43 & 700 & 1.2 & 5.5 & -4.26 \\
VXR44   & -0.3867\sups{g} & -3.95      & 270 & 2.0 & 1.6 & -4.80 & 240 & 1.8 & 1.5 & -4.82 & 310 & 2.3 & 1.9 & -4.72 \\
VXR45A & -1.9420               & -3.62      & 630 & 2.1 & 4.9 & -4.31 & 530 & 1.8 & 4.2 & -4.38 & 510 & 1.7 & 4.0 & -4.40 \\
VXR50A & -1.6840\sups{g} & -3.32      & 740 & 2.5 & 6.0 & -4.22 & 480 & 1.6 & 3.8 & -4.42 & 650 & 2.1 & 5.0 & -4.30 \\
VXR52   & -0.3509\sups{g} & -4.16      & 220 & 1.8 & 1.3 & -4.89 & 150 & 1.2 & 0.9 & -5.05 & 220 & 1.8 & 1.3 & -4.89 \\
VXR62A & -1.6202               & -3.30      & 570 & 1.6 & 4.5 & -4.35 & 360 & 1.0 & 2.8 & -4.55 & 550 & 1.5 & 4.2 & -4.38 \\
VXR66   & -0.4982\sups{g} & -4.59      & 120 & 0.8 & 0.8 & -5.10 &  80 & 0.5 & 0.5 & -5.30 & 130 & 0.8 & 0.8 & -5.10 \\
VXR67A & -0.7075\sups{g} & -3.45     & 490 & 1.3 & 4.0 & -4.40 & 440 & 1.2 & 3.7 & -4.43 & 440 & 1.1 & 3.4 & -4.47 \\
VXR69A & -1.0840\sups{g} & -3.48     & 510 & 1.7 & 4.0 & -4.40 & 380 & 1.3 & 3.0 & -4.52 & 460 & 1.5 & 3.5 & -4.46 \\
VXR70   & -0.6171                & -3.77     & 290 & 1.3 & 2.2 & -4.66 & 190 & 0.9 & 1.5 & -4.82 & 290 & 1.3 & 2.2 & -4.66 \\
VXR72   & -0.6970                & -3.21     & 450 & 1.7 & 3.5 & -4.46 & 330 & 1.2 & 2.5 & -4.60 & 380 & 1.4 & 2.9 & -4.54 \\
VXR76A & -0.6815               & -3.11     & 600 & 1.1 & 4.9 & -4.31 & 460 & 0.8 & 3.5 & -4.46 & 500 & 0.9 & 4.0 & -4.40 \\
VXR77A & -0.7799               & -3.77     & 310 & 1.9 & 2.2 & -4.66 & 240 & 1.4 & 1.6 & -4.80 & 360 & 2.1 & 2.4 & -4.62 \\
VXR78   & -1.1189\sups{g} & -3.14     & 650 & 3.0 & 4.8 & -4.32 & 380 & 1.8 & 2.9 & -4.54 & 510 & 2.3 & 3.7 & -4.43 \\
VXR80A & -1.9622\sups{g} & -3.29     & 780 & 2.0 & 6.2 & -4.21 & 500 & 1.3 & 4.0 & -4.40 & 700 & 1.8 & 5.6 & -4.25 \\
\hline
\end{tabular}
\end{minipage}
\end{center}
\end{table*}
\end{landscape}

\begin{landscape}
\begin{table*}
\begin{center}
\begin{minipage}{240mm}
\caption{Chromospheric IRT emission of the single members observed in IC 2602. The Columns are the same as in Table~\ref{irt2391chromo}, except that the X-ray emission is taken
from Stauffer et al.\ (1997).} \label{irt2602chromo}
\scriptsize
\begin{tabular}{lcccccccccccccc}
\hline
Star & Log(N\subs{R}) & Log(L\subs{X}/L\subs{bol}) & W$^{\rm em}_{\rm 8542}$ & F$^{\prime}_{\rm 8542}$ (x10\sups{3}) & R$^{\prime}_{\rm 8542}$ & Log(R$^{\prime}_{\rm 8542})$ & W$^{\rm em}_{\rm 8498}$ & F$^{\prime}_{\rm 8498}$ (x10\sups{3}) & R$^{\prime}_{\rm 8498}$ & Log(R$^{\prime}_{\rm 8498}$) & W$^{\rm em}_{\rm 8662}$ & F$^{\prime}_{\rm 8662}$ (x10\sups{3}) & R$^{\prime}_{\rm 8662}$ & Log(R$^{\prime}_{\rm 8662}$) \\
Name &                &                            & (m\AA)                  & (Wm\sups{-2})                         & (x10\sups{-5})          &                              & (m\AA)                  & (Wm\sups{-2})                         & (x10\sups{-5})          &                              & (m\AA)                  & (Wm\sups{-2})                         & (x10\sups{-5})          &                              \\
\hline
R1     & -0.6720\sups{g}  &                 & 540 & 1.9 & 4.2 & -4.38 & 350 & 1.3 & 2.9 & -4.54 & 420 & 1.5 & 3.3 & -4.48 \\
R7     & -0.1719\sups{g}  & -4.70       & 120 & 1.0 & 0.7 & -5.15 &  70  & 0.6 & 0.4 & -5.40 &  70  & 0.6 & 0.4 & -5.40 \\
R8A   & -0.8175\sups{g}  &                 & 320 & 1.8 & 2.2 & -4.66 & 180 & 1.0 & 1.2 & -4.92 & 250 & 1.4 & 1.7 & -4.77 \\
R10   & -1.0748\sups{g}  &                 & 470 & 0.9 & 3.8 & -4.42 & 370 & 0.7 & 3.0 & -4.52 & 460 & 0.9 & 3.8 & -4.42 \\
R14   & -0.8475\sups{g}  & -3.40       & 570 & 2.2 & 4.5 & -4.35 & 410 & 1.6 & 3.3 & -4.48 & 500 & 1.9 & 3.9 & -4.41 \\
R15A & -0.7807                & -2.92       & 730 & 2.0 & 6.0 & -4.22 & 570 & 1.5 & 4.5 & -4.35 & 540 & 1.4 & 4.2 & -4.38 \\
R29   & -1.0274                & -3.31       & 650 & 1.4 & 5.3 & -4.28 & 460 & 1.0 & 3.8 & -4.42 & 520 & 1.1 & 4.2 & -4.38 \\
R35   &  0.4063\sups{g}  & -4.09       & 390 & 2.0 & 2.7 & -4.57 & 260 & 1.4 & 1.9 & -4.72 & 290 & 1.5 & 2.1 & -4.68 \\
R42C & -0.7477\sups{g} &                 & 580 & 2.2 & 4.4 & -4.36 & 430 & 1.7 & 3.4 & -4.47 & 450 & 1.7 & 3.4 & -4.47 \\
R43   & -1.4513                & -3.05       & 940 & 2.4 & 7.8 & -4.11 & 600 & 1.5 & 4.9 & -4.31 & 770 & 1.9 & 6.2 & -4.21 \\
R45A & -0.6032\sups{g} & -4.29       & 370 & 1.9 & 2.7 & -4.57 & 250 & 1.3 & 1.8 & -4.74 & 270 & 1.4 & 2.0 & -4.70 \\
R52   & -1.3977                & -3.42       & 790 & 2.0 & 6.6 & -4.18 & 450 & 1.1 & 3.6 & -4.44 & 670 & 1.6 & 5.3 & -4.28 \\
R58   & -1.2294                & -3.22       & 720 & 3.5 & 5.2 & -4.28 & 410 & 2.0 & 3.0 & -4.52 & 600 & 2.9 & 4.3 & -4.37 \\
R59   & -1.1527                & -3.13       & 780 & 2.4 & 6.4 & -4.19 & 520 & 1.6 & 4.3 & -4.37 & 630 & 1.9 & 5.1 & -4.29 \\
R66   & -0.5386                & -3.76       & 370 & 1.5 & 2.7 & -4.57 & 250 & 1.0 & 1.8 & -4.74 & 280 & 1.1 & 2.0 & -4.70 \\
R68   & -1.3063                & -3.03       & 870 & 2.2 & 7.0 & -4.15 & 540 & 1.4 & 4.5 & -4.35 & 640 & 1.6 & 5.1 & -4.29 \\
R70   & -0.4453                & -4.44       & 280 & 1.5 & 2.1 & -4.68 & 210 & 1.1 & 1.5 & -4.82 & 240 & 1.2 & 1.6 & -4.80 \\
R72   & -0.9377                & -3.01       & 770 & 3.7 & 5.7 & -4.24 & 500 & 2.4 & 3.7 & -4.43 & 620 & 2.9 & 4.5 & -4.35 \\
R79   & -0.2684\sups{g} & -4.38       & 160 & 1.2 & 1.0 & -5.00 & 120 & 0.9 & 0.7 & -5.15 & 100 & 0.8 & 0.6 & -5.22 \\
R80   & -0.5173               & -3.69        & 330 & 1.0 & 2.7 & -4.57 & 220 & 0.7 & 1.9 & -4.72 & 240 & 0.7 & 1.9 & -4.72 \\
R83   & -0.6796               & -3.55        & 520 & 2.4 & 3.9 & -4.41 & 370 & 1.7 & 2.8 & -4.55 & 440 & 2.0 & 3.2 & -4.49 \\
R85   & -0.5466\sups{g} & -4.74       & 150 & 1.0 & 1.0 & -5.00 &  80  & 0.5 & 0.5 & -5.30 & 110 & 0.7 & 0.7 & -5.15 \\
R88A & -2.0748              & -3.54        & 740 & 1.2 & 5.9 & -4.23 & 530 & 0.8 & 3.9 & -4.41 & 520 & 0.8 & 3.9 & -4.41 \\
R89   & -0.7088               & -3.54       & 570 & 0.9 & 4.4 & -4.36 & 430  & 0.7 & 3.4 & -4.47 & 450 & 0.7 & 3.4 & -4.47 \\
R92   & -0.7777               & -3.72        & 430 & 1.9 & 3.2 & -4.49 & 270 & 1.2 & 2.0 & -4.70 & 300 & 1.3 & 2.2 & -4.66 \\
R95A & -1.2292              & -2.96        & 690 & 2.2 & 5.5 & -4.26 & 500 & 1.6 & 4.0 & -4.40 & 570 & 1.8 & 4.5 & -4.35 \\
R96   & -1.1313              & -3.19         & 470 & 0.7 & 3.5 & -4.46 & 400 & 0.6 & 3.0 & -4.52 & 480 & 0.7 & 3.5 & -4.46 \\
W79  & -0.4866              & $<$ -4.31 & 300 & 1.2 & 2.3 & -4.64 & 220 & 0.9 & 1.7 & -4.77 & 220 & 0.9 & 1.7 & -4.77 \\
\hline
\end{tabular}
\end{minipage}
\end{center}
\end{table*}
\end{landscape}